\documentclass[secnumarabic,amssymb,nofootinbib. aps, prd,onecolumn]{revtex4-2}% nobibnotes,
\usepackage{graphicx,epsfig,youngtab}

\pdfoutput=1
\usepackage{xcolor}
\usepackage{bbold}
\usepackage{amsmath}
\usepackage{hyperref}
\usepackage{amsfonts}
\expandafter\let\csname equation*\endcsname\relax 
\expandafter\let\csname endequation*\endcsname\relax 
\usepackage{bm}
\usepackage{lmodern}
\usepackage{natbib}

\def\x {\bm{x}}

\def\d {{\rm{d}}}

\def\k {\bm{k}}
\def\x {\bm{x}}

\newcommand{\p}{_{{\text{\tiny$\|$}}}}% ||

\newcommand{\n}{{\bf \hat{n}}}
\newcommand{\HH}{\mathcal{H}\,}

\newcommand{\two}{^{\text{\tiny \color{green}{({{2}})}}}}
\newcommand{\one}{^{\text{\tiny \color{red}{({{1}})}}}}

\newcommand{\<}{\langle}
\renewcommand{\>}{\rangle}

\def\jnlref#1{{\rm#1}}
\newcommand\apjl{The Astrophysical Journal Letters}
\def\mnras{\jnlref{MNRAS}}
\def\aapr{\jnlref{A\&A~Rev.}}
\def\pasp{\jnlref{PASP}}
\def\aap{\jnlref{A\&A}}

\newcommand{\averageA}[1]{\left\langle #1 \right\rangle_{\Omega}}

\begin{document}
   % \title{\boldmath Fitting the FLRW spacetime to observation on ultra-small redshifts \\  and the infered Hubble rates: Paper II} 
        % \title{\boldmath Cosmological fitting problem and the Hubble tension: Paper II} 
        % \title{\boldmath Area distance in a lumpy universe at ultra-small redshift limit} 
          %\title{\boldmath Area distance in the neighbourhood of an observer} 
         \title{\boldmath The art of building a smooth cosmic distance ladder in a perturbed universe} 

\author{Obinna Umeh }
%\affiliation {Institute of Cosmology \& Gravitation, University of Portsmouth, Portsmouth PO1 3FX, United Kingdom}
\affiliation {${}^{1}$Institute of Cosmology \& Gravitation, University of Portsmouth, Portsmouth PO1 3FX, United Kingdom
\\
${}^{2}$Department of Physics, University of the Western Cape,
Cape Town 7535, South Africa 
\\
${}^{3}$Department of Mathematics and Applied Mathematics, University of Cape Town, Rondebosch 7701, South Africa}
%\email{
%obinna.umeh@port.ac.uk
%}
\email{obinna.umeh@port.ac.uk}
\date{\today}

\begin{abstract}

How does a smooth cosmic distance ladder emerge from observations made from a single location in a lumpy Universe? 
Distances to Type Ia supernovae  in the Hubble flow are anchored on local distance measurements to sources that are very nearby. We described how this configuration could be built in a perturbed universe where lumpiness is described as small perturbations on top of a flat Friedmann-Lemaıtre Robertson-Walker  spacetime. 
We show that there is a non-negligible modification (about 11\%) to the background Friedmann-Lemaıtre Robertson-Walker area distance due to the presence of inhomogeneities in the immediate neighbourhood of an observer. We find that the modification is sourced by the electric part of the Weyl tensor indicating a tidal deformation of the local spacetime of the observer.  
We show in detail how it could impact the calibration of the Type Ia supernova absolute magnitude in the Hubble flow. We show that it {could potentially} resolve the Type Ia supernova absolute magnitude and Hubble tensions simultaneously without the need for early or late dark energy.

\end{abstract}

\maketitle
\DeclareGraphicsRule{.wmf}{bmp}{jpg}{}{}
%\maketitle

\tableofcontents
\maketitle
%\newpage

\section{Introduction}

The observable patch of the universe from earth has structures of different sizes and shapes interacting gravitationally among themselves and with the observer~\cite{Trujillo:2005xf,Hahn:2006mk,Libeskind:2017tun}. The apparent luminosities from these sources are interpreted assuming exact cosmological principle, i.e isotropic and homogeneous Friedmann-Lemaıtre Robertson-Walker (FLRW) model~\cite{2010fimv.book..267S,Thepsuriya:2014zda,Riess:2016jrr,Freedman:2021ahq}. The symmetry of the FLRW spacetime does not allow such a local inhomogeneous distribution of structures as seen by an observer. Yet the standard model of cosmology assumes it~\cite{Aghanim:2018eyx,Freedman:2019jwv,Riess:2021jrx}. 
The key justification for this was provided by Weinberg in 1976. Weinberg showed that the impact of gravitational deflection due to the inhomogeneities on the apparent luminosities of sources is the same on average to the universe where the inhomogeneities were spread out homogeneously through space~\cite{1976ApJ...208L...1W}. The proof assumes the following: that the cosmological principle holds~\cite{Ellis:1998ha}, that the time and position in the FLRW and the inhomogeneous spacetime are synchronised~\cite{Breton:2020puw}, that  there is no strong  gravitational lensing events or caustics as photons traverse over-densities along the line of sight~\cite{Ellis:1998qga,Ellis:2018led}  and that the redshift is a monotonic function on all scales even in the presence of structures.

Our target here is to explore in detail using the cosmological perturbation theory the assumption that the redshift is a monotonic function on all scales by calculating the monopole of the area distance on a constant redshift surface. This is very crucial because the local distance ladder is calibrated using distance measurement to nearby sources~\cite{Freedman:2019jwv,Riess:2021jrx}. It is well-known that distance to those sources is not a simple function redshift~\cite{Davis:2010jq,Carr:2021lcj,Peterson:2021hel}.
Several works have studied the luminosity distance (distance modulus) in perturbation theory before~\cite{Umeh:2012pn,Umeh:2013UCT,Umeh:2014ana,Ben-Dayan:2012uam,Ben-Dayan:2012lcv,BenDayan:2012ct,Ben-Dayan:2014swa,Fanizza:2013doa,Fanizza:2015swa}, the result from these efforts show that the background FLRW spacetime is a good approximation of the area distance in the observed universe~\cite{Clarkson:2014pda,Kaiser:2015iia}. 
Independent studies based on general relativistic N-body simulation reached a similar conclusion \cite{Adamek:2018rru,Tian:2020qnm,Breton:2020puw}.  
However, these studies focused on the following redshift limit $0.023\le z \le1100$~\cite{Clarkson:2014pda,Bonvin:2015uha,Adamek:2018rru,Adamek:2018rru,Tian:2020qnm,Breton:2020puw}. The focus on this redshift range was motivated by the fact that the SHoES collaboration and the Carnegie-Chicago Hubble Program (CCHP) truncated the sample of the peak magnitudes of the SNIa at $z\ge 0.02$~\cite{Riess:2016jrr,Freedman:2019jwv}.  
On the surface, this appears to be sufficiently motivated but on closer examination, it becomes clear that the determination of the Hubble rate using the peak magnitudes of Type Ia supernova (SNIa) within this redshift range ($0.023\le z\le 0.5$) relies crucially on the geometric distance measurement of a set of anchors that are very nearby~\cite{Riess:2016jrr,Freedman:2019jwv}. Essentially, the calibrated absolute magnitude of the SNIa in the Hubble flow, could depend on the distance measurement to the anchors in  the $ z\le 0.01$ redshift range~\cite{Lombriser:2019ahl}. {This is very important because the absolute magnitude of the  SNIa in the Hubble flow is simply the $\log_{10}$ of the distance  measured at 1Mpc.}  Such a dependence could manifest as a tension in absolute magnitude of the SNIa~\cite{Camarena:2019rmj,Camarena:2021jlr,Efstathiou:2021ocp} if the anchors live in  a local over-density with a non-trivial curvature. 
%This topic was explored in  but from the perspective that the anchors live in a local under-density. We do not make such an assumption here, we simply make use of the general relativistic perturbation theory to show that the anchors live in an over-density.

To build a consistent cosmic distance ladder in a perturbed universe that captures the interplay between the geometric distance measurements to a set of anchors and the SNIa in  ($0.023\le z\le 0.5$) redshift range, we derive a very concise expression for the area distance that includes general relativistic corrections in a perturbed FLRW spacetime. The full general relativistic expression for the area distance within the cosmological perturbation theory is humongous, it covers more than a page when written down~\cite{Umeh:2012pn,Umeh:2014ana}. This usually involves perturbing the Sach's equation for the area distance~\cite{Umeh:2012pn,Umeh:2014ana} or perturbing the Jacobi map \cite{Fanizza:2015swa,Fanizza:2013doa} or using the light-cone formalism to decompose the Jacobi map \cite{Ben-Dayan:2012uam,Ben-Dayan:2012lcv,BenDayan:2012ct}. The formalism we present here is much simpler and points straight to the key terms.
Using this expression, we show that at very small redshifts (limit where the redshift goes to zero), the monopole of the distance-redshift relation is determined by the rate of shear deformations of the local spacetime. Distance to the nearby anchors is not given by the FLRW spacetime since the rate of shear deformation tensor vanishes in an FLRW spacetime.
We show that for a consistently calibrated zero-point of the distance modulus, the SHoES and CCHP teams should infer the same Hubble rates and the Hubble rate inferred by both teams corresponds to global volume expansion.
Finally, by fitting the Alcock-Paczynski parameters ~\cite{Alcock:1979mp} based on the FLRW spacetime to corresponding perturbed expression, we show that the Hubble rate determined this way is sensitive to the local environment. {The key result on area distance is given in equation \eqref{eq:averageareadistance} and the possible resolution of the supernova absolute magnitude tension is given in equation \eqref{eq:Mdiff}. }

The rest of the paper is organised as follows: We give a very concise derivation of the area distance that isolates cleanly the effect of radial and tangential distortions of the source position in section \ref{sec:distance_modulus}.  We show how the tidal effects contribute to the monopole of the area distance at a very small redshift in sub-section \ref{sec:tidaldeformation}.
We discuss the consequences of the tidal deformation in section \ref{sec:cosmography}. In sub-section \ref{sec:correctHO} , we discuss the calibration of the SNIa peak magnitude using the TRGB and in sub-section \ref{sec:SHoES} we discuss the calibration using cepheids. The discussion on the SNIa absolute magnitude tension is provided in sub-section \ref{sec:SN1Amagtension}. In sub-section \ref{sec:BAO}, we illustrate how the Alcock-Paczynski parameters might be interpreted given a perturbed expression for the area distance and the Hubble rate.  We conclude in section \ref{sec:conc}. The details of the derivation of the area distance in a perturbed universe are given in the Appendix \ref{deviationvector}.
\\
{\bf{Cosmology}}: We adopt the following values for the cosmological parameters of the standard model \cite{Aghanim:2018eyx}: the dimensionless Hubble parameter, $h = 0.674$, baryon density parameter, $\Omega_{\rm b} = 0.0493$,  dark matter density parameter, $\Omega_{\rm{cdm}} = 0.264$,  matter density parameter, 
$\Omega_{\rm m} = \Omega_{\rm{cdm}} + \Omega_{\rm b}$,  spectral index, $n_{\rm s} = 0.9608$,  and the amplitude of the primordial perturbation, $A_{\rm s} = 2.198 \times 10^{-9}$. 
The small English alphabets from $a-e$ denotes the full spacetime indices, where $i$ and $j$ denote the spatial indices.

\section{Area distance in a Universe with structures}\label{sec:distance_modulus}

Sources are detected at their image positions (apparent position), $x^a_{I}$, they are related to their
 physical position  ${x}^a_{\rm{s}}$ via a general coordinate transformation:
\begin{eqnarray}\label{eq:generalcoord}
{x}^a_{\rm{s}}(x')&=& x^a_{I}+ \left(\frac{\partial x^a}{\partial x^{'b}}\right)_{I}\left(x'^b-x^{b}_{I}\right)
+ \mathcal{O}\left(\bm{x}'-\bm{x}_{I}\right)^2 \,,
\end{eqnarray}
%where $x^a_I = x'^a(\lambda_s)$ is the image/apparent position that an observer from a single location will  infer about the source at an affine parameter distance $\lambda_s$. %In polar coordinates in 4-dimension,   
In the Geometric optics limit, we can describe light rays which carry distance information as affinely parameterised null geodesics ${k}^b{\nabla}_b{ k}^a=0$, where $k^a$ is the photon  tangent vector to the null geodesic emanating from a source at position $x^a_s$ and converging at the observer at $x^a_o$. For points sources, $x^a_s$ and  $x^a_o$ constitute a conjugate pair. Here, ${\nabla}$ is the spacetime covariant derivative and $k^a$ satisfies the Eikonal equation ${k}_a{k}^a=0$. 
This  photon trajectory traces a curve which is parametrised by the affine parameter $\lambda$ according
\begin{eqnarray}\label{eq:positioneqn}
k^a = \frac{\d x^a}{\d\lambda}\,. %=k^b\partial_b \,x^a\,.
\end{eqnarray}
Given a comoving observer with a time-like 4-velocity, $u^a$ ($u^au_a=-1$),  we  decompose $k^a$ into parallel and orthogonal  components  with respect to  $u^a$: ${k}^a = E({u}^a\pm{n}^a)$, where $E$ is the photon energy, $E= -u_ak^a$ and $n^a$ is a space-like  line of sight (LoS) direction vector ($n^an_a = 1$).  
{In this set-up,  we consider a one-parameter family geodesics, among the one-parameter family, we choose a ``central ray" as a ``unique ray" and calculate deviation of the infinitesimally close nearby geodesics. In general the deviation is attributed to the  effect of gravity~\cite{1984ucp..book.....W}. The result we get at the end is independent of this choice since what is important is the difference~\cite{1966ApJ...143..379K,Lewis:2006fu}.}
Treating $x^a_{I}(\lambda)$ as the trajectory of the central ray, we define a difference/connecting/deviation vector as  the difference between the inferred apparent position and the physical position of the source at the same affine parameter distance $\lambda$:  $ x'^a(\lambda)- x^a_{I}(\lambda) \equiv \xi'^a $. Putting this into equation \eqref{eq:generalcoord}  gives
\begin{equation}\label{eq:jacobi_def}
 \xi^a =  \mathcal{J}^{a}{}_{b}\xi'^b
 + \mathcal{O}({\xi'^b})^{2}\,,
 \end{equation}
where we have introduced the Jacobian, $\mathcal{J}^{a}{}_{b} =  \partial x^{a}/\partial x^b_{I} = \delta_{ab} + \mathcal{D}_{ab}$, with $\mathcal{D}_{ab}=\partial_{ b} \xi_a$. We work in  orthonormal basis, $\mathbf{e}^a$, such that the rest frame of the observer has a time-like orthonormal basis vector ${e}^a{}_0 = {u}^a$ and the radial vector has a orthonormal basis vector ${e}^a{}_1 = {n}^a$. We denote the two angular orthonormal basis vector as ${e}^a{}_A$, where, $A={2,3}$. The  screen space metric(the two space orthogonal to the geodesics) is denoted by {$N^{ab}= {g}^{ab}+{u}^a{u}^b-{n}^a{n}^b $ and $g^{ab}$ is the metric of the full spacetime.} It is used to project the spacetime indices to the orthonormal tetrads indices on the screen space; $N_{ab} e^a{}_{A} e^b{}_B = \eta_{AB}$ and $ e^a{}_{A} = N^a{}_b e^b{}_{A}$. 

{For the irreducible decomposition of the Jacobian, we follow the formalism introduced in \cite{Schmidt:2012ne,Jeong:2011as}. The computation of the area distance within this formalism requires  only the screen space projected components of the Jacobian. To obtain  this part, it is enough to decompose just the spatial component of the deviation vector $\xi^i$ with respect to $n^i$: $\xi^i=\xi_{\p} n^{i}  +  \xi_{\bot}^i$, where $\xi_{\p}  = \xi_i n^i$ is the component parallel to $n^{i}$ and $\xi_{\bot A} = e^i{}_{A} \xi_{i}$ is the angular or the screen space projected part. Similarly, the Jacobian  can also be decomposed as
\begin{eqnarray}
\mathcal{D}_{ij} & =&  n_{i}n_{j} \mathcal{D}{\p}{\p}+ 2n_{(i}e_{j)}{}^A\mathcal{D}_{\bot A}{ \p}  + e_{i}{}^Ae_{j}{}^B \mathcal{D}_{\bot AB} \,,
\end{eqnarray}
%+ \partial_{\bot a} n_{b} \xi\p 
where $\mathcal{D}{\p}{ \p }=  \mathcal{D}_{i j } n^{i} n^{j}\,,$ $ \mathcal{D}_{\bot A_1}{ \p}  =n^{i} e^{j}{}_{A_1}  \mathcal{D}_{ ij}$  and $\mathcal{D}_{ \bot A_{1}A_{2}} = e^{i}{}_{A_1}  e^{j}{}_{A_2}\mathcal{D}_{ ij}$.  
$n^i \partial_i n^j =0$ vanishes on the Minkowski spacetime(or on the conformal FLRW spacetime). The angular part of the Jacobian, $ \mathcal{D}_{\bot AB} $  may be decomposed further into irreducible parts:
 \begin{eqnarray}
 \mathcal{D}_{ \bot A B}& =& e^{i}_{A} e^{j}_{B}\mathcal{D}_{ ij} \,,
 \\
 &=& e^{i}_{A} e^{j}_{B}\left[ \frac{1}{2}\bar{\theta}\xi_{\p} \delta_{ij} +  \partial_{\bot i} \xi_{\bot j}\right] \,,
 \\ 
 &=& e^{i}_{A} e^{j}_{B}\left[ \frac{1}{2}\bar{\theta}\xi_{\p} \delta_{ij} + N_{ij} \kappa + \varepsilon_{ij}\omega + \gamma_{ij}\right]
 =
 \frac{1}{2}\mathcal{D}\delta_{AB} + \gamma_{\<AB\>} + \varepsilon_{AB}\omega\,,% \omega_{[AB]} .
 \label{decompJacobi0}
\end{eqnarray}
where $\varepsilon_{AB}$  is the Levi-Civita tensor on the screen space, we  made use of the decomposition of $\xi_i$ with respect to $n^i$ in the second equality, $\bar{\theta}$ is the divergence of $n^i$ on the conformal FLRW spacetime and we introduced the following terms in the last equality
\begin{eqnarray}
  && \mathcal{D}_{\bot}=\mathcal{D}_{A}{}^A \,, \qquad 
  \gamma_{\< AB\>}= \mathcal{D}_{{\bot}( AB)} - \frac{1}{2}N_{AB}\mathcal{D}_{{\bot}C}{}^C 
   \,, \qquad
   \omega = \frac{1}{2}\varepsilon^{AB}\mathcal{D}_{{\bot}AB} \,.
   \end{eqnarray}
where  $\mathcal{D}$ is the trace of $\mathcal{D}_{AB}$, it describes the isotropic distortion of the image of the source due to inhomogeneity. Its irreducible form is given by 
\begin{eqnarray}
 \mathcal{D}_{\bot}= \partial_{\bot A} n^{A}_{\bot} \xi_{||}-2\kappa= \frac{2}{ r} \xi_{||}-2\kappa\,,
  \end{eqnarray} 
 where $\kappa$ is the weak gravitational lensing convergence,  $\xi_{||} \partial_{\bot A} n^{A}_{\bot} $ is a general relativistic  correction to the weak gravitational lensing convergence. The importance of this term ($\xi_{\||}$ was first recognised in \cite{Ellis:1998qga}, where it was pointed out that it describes the effect of the radial lensing. It has a natural interpretation in redshift space where its describes the Doppler lensing effects, as we shall see the leading order part describes the parallax effect which usually exploited to estimate geometric distances to nearby sources. When we say  parallax effect, we mean the displacements in the source position due to the relative motion of the observer with respective to the local over-density.  On the conformal FLRW background, $\bar{\theta}$ is given by $\bar{\theta} = \partial_{\bot A} n^{A}_{\bot} = {2}/{r}$.    Similarly,   $\partial_{\bot [A} n_{\bot B]} = 0$  and $ \partial_{\bot \<A}n_{\bot B\>} = 0$ on the conformal FLRW background.}
 Furthermore, $\gamma_{\<AB\>}$ is the shear, it corresponds to the the trace-free part of $\mathcal{D}_{\bot AB}$, it describes the anisotropic stretching of initial circular image and $\omega$ is the twist, it corresponds to anti-symmetric part of $\mathcal{D}_{\bot AB}$.
These observables are defined in terms of the deviation vector:
   \begin{eqnarray}\label{eq:irreduciabledecomp}
\kappa &=&- \frac{1}{2}N^{AB} \nabla_{\bot A} \xi_{\bot  B}\,,
   \qquad
   \gamma_{AB} =\nabla_{\bot \< A} \xi_{\bot B\>}\,,
   \qquad
   \omega =  \varepsilon^{AB}\nabla_{\bot [A}\xi_{ \bot B]}\,.
   \end{eqnarray}
The area distance, $d_A$,  is related to the area  element of the screen space(area of the wave-front advancing  towards from the past)  according to 
$
{\d} A = \sqrt{{N} }\,{ \d}\Omega= d_A^2 {\d}\Omega\,,
$
where ${N}$ is the determinant of the induced physical metric on the screen space.  It is related to the  metric of the celestial sky, $\Omega_{AB}$,   according to the general metric transformation law
\begin{eqnarray}\label{eq:transgab}
N_{AB} =\frac{\partial {x}^C}{\partial x'^A}\frac{\partial {x}^D}{\partial x'^B}\, \, \Omega_{CD} \,.
\end{eqnarray}
The determinant of a product of matrices is equal to the product of the determinants of each matrix: $\text{Det}[\bm{N}]= \text{Det}[{\bm{\Omega}}]\text{Det}[{\bf{\mathcal{{J}}}}]^2$, therefore, we can  use the Cayley–Hamilton theorem to evaluate the  determinant of $\text{Det}[{\bf{\mathcal{{J}}}}]$ since $ \text{Det}[{\bm{\Omega}}]=1$
\begin{eqnarray}\label{eq:defareadistance}
\text{Det}[ \mathcal{J}]&=&\text{Det}[\bar{ \mathcal{J}}]\bigg[
I+ \mathcal{D}^i{}_i +\frac{1}{2}\left(\mathcal{D}^i{}_i\mathcal{D}^j{}_j - \mathcal{D}^{ij}\mathcal{D}_{ij}\right)+ \mathcal{O}(\epsilon^3)\bigg]\,,
\end{eqnarray} 
where $\text{Det}(\bar{ \mathcal{J}}_{ij})$ is related to the area distance on the background spacetime according to $\bar{d}_A^2=\text{Det}(\bar{ \mathcal{J}}_{ij})$ and{$\varepsilon$ is an infinitesimally small  cosmological perturbation theory parameter}.  On the full spacetime ${d}_A^2=\text{Det}({ \mathcal{J}}_{ij})$ and taking the square root gives
\begin{eqnarray}\label{eq:areadistance}
d_A&=&\bar{d}_A\bigg[1+\frac{ \xi{\p}}{\bar{d}_{A}} - \kappa-\frac{1}{4}\left( \gamma_{ij} \gamma^{ij}-\omega_{ij}\omega^{ij}\right) +\mathcal{O}(\epsilon^3)\bigg]\,.
\end{eqnarray} 
{An observer on Earth has a non-vanishing peculiar motion  with respect to the cosmic rest frame. Most telescopes, automatically corrects for the the impact of the relative velocity between the Earth and the sun so that observables are given in the heliocentric frame~\cite{Courteau:1999aa}. 
Let's denote the non-vanishing relative velocity between the heliocentric frame and the cosmic rest frame with $v^a_{\rm{sun-CRF}}$. This  contribution of the heliocentric relative velocity distorts the solid angle. %Using the definition of the area distance given in equation \eqref{eq:areadistance} \cite{Kaiser:2014jca,Kaiser:2015ada}. 
The distortion of the solid angle due to $v^a_{\rm{sun-CRF}}$ is usually implemented by boosting $d_A$ in equation \eqref{eq:areadistance} to the heliocentric frame which has a non-vanishing peculiar velocity. Let $\tilde{u}^a$ denote the four-velocity of the observer in this frame.  The relationship between $\tilde{u}^a$  and $u^a$ is given by~\cite{Maartens:1998xg}
 \begin{eqnarray}\label{eq:4velocity_def}
\tilde{u}^a=\gamma(u^a+v^a_{\rm{sun-CRF}}),~~~\text{where}~~~\gamma=\frac{1}{\sqrt{1-v^2_{\rm{sun-CRF}}}},~~~v^a_{\rm{sun-CRF}}u_a=0\,,
 \end{eqnarray}
The solid angles in both frames are related according to~\cite{Davis:2010jq,Bonvin:2005ps}: 
 \begin{eqnarray}
 \frac{\d \Omega}{\d \tilde{\Omega}}&=&\left(\frac{{E}}{\tilde{E}}\right)^2 
 =\gamma^2(1-n^av_{a\rm{sun-CRF}})^2\,,
 \end{eqnarray}
 where  ${\tilde{E}}/{{E}}=\gamma(1-n^av_{a\rm{sun-CRF}})$ \cite{Challinor:2000as}. Then the  infinitesimal change in the area becomes $ {\d} A = \tilde{d}_A^2\big| {\d}\Omega/\d\tilde{\Omega}\big| \d\tilde{\Omega}\,,$
hence, the relationship  between the area distance in both frames becomes
 \begin{eqnarray}
  \tilde{d}_A=\frac{\tilde{E}}{{E}}\bigg|_{o}  d_A \approx \left[1- n^iv_{i{\rm{sun-CRF}}} + n^i n^jv_{i{\rm{sun-CRF}}} v_{j{\rm{sun-CRF}}} + \frac{1}{2} v^{a}_{{\rm{sun-CRF}}}v_{a {\rm{sun-CRF}}}\right] d_A \,.
  \label{eq:da_velocity}
  \end{eqnarray}
As we shall describe in detail later and elsewhere using the Jacobi field formalism, this additional contribution to the area distance due to the heliocentric observer peculiar velocity are not describable within the cosmological perturbation theory on an FLRW background without the addition of the higher multipoles of the peculiar velocity field. Please see~\cite{Macaulay:2010ji} for details on this topic. More discussion on this is provided in sub-section \ref{sec:redshiftperturbation}.  

For consistency or sanity check, we can expand equation \eqref{eq:da_velocity} up to linear order  in velocity to find $  \tilde{d}_A=\bar{d}_{A}(1- n^iv_{i \rm{sun-CRF}}+ { \xi{\p}}/{\bar{d}_{A}} + \kappa|_{\rm{linear}} )$. This helps to see immediately that  all the general relativistic corrections described in \cite{Bonvin:2005ps,Umeh:2012pn,Umeh:2014ana} at linear order are contained in $n^iv_{oi}+ { \xi{\p}}/{\bar{d}_{A}}$.  
Our main focus here is to examine this term at second order. 
The impact of the observer peculiar velocity on the distance  measurement  was discussed in \cite{Kaiser:2014jca,Kaiser:2015ada}. 

}

\subsection{General relativistic correction to the area distance on constant redshift surfaces}\label{sec:perturbationtheory}

We work in Conformal Newtonian  gauge on  a flat FLRW background spacetime for simplicity:
\begin{eqnarray}\label{eq:metric}
\d s^2 &=& a^2\big\{-\big[1 + 2\Phi \big]\d \eta^2 + \big[1-2 \Psi\big]\delta_{ij} \d x^{i}\d x^{j}\big\}\,,
\end{eqnarray}
where $\delta_{ij}$ is the metric of the spatial section of the Minkowski spacetime, $\Phi$  and  $\Psi$ are  the Newtonian and curvature  potential respectively. In the limit of vanishing anisotropic tensor in General relativity, $\Phi = \Psi$. 
The components of $u^a$ perturbed up to second order in  perturbation theory are given by
\begin{eqnarray}
u^0&=&1 - \Phi +  \frac{3}{2} \Phi^2 -  \frac{1}{2}\Phi\two + \frac{1}{2}{\partial}_{i}v{\partial}^{i}v\,,
\\ 
u^i&=&{\partial}^{i}v + \frac{1}{2}
v^{i}{}\two + \frac{1}{2} {\partial}^{i}v\two\,,
\end{eqnarray}
where $v^i$ is the peculiar velocity,   $\partial_i$ is the spatial derivative  with respective to the background spacetime and $v$ is the peculiar velocity potential. 
The metric perturbation in equation \eqref{eq:metric}  remains valid except in the vicinity of a very strong gravitational field source such as a black hole~\cite{Ishibashi:2005sj}. In general, 
$\Phi$ satisfies
\begin{eqnarray}
 |\Phi| \ll 1 \,, \qquad 
 \left| {\Phi'} \right|^2   
    \ll  \partial^i\Phi \partial_i \Phi \,, \qquad 
 (\partial^i \Phi \partial_i \Phi)^2 \ll \partial^i\partial^j\Phi \partial_i\partial_j\Phi \,.  
\label{condi:Newtonian}  
\end{eqnarray}
{where $'$ is the partial derivative with respect to the conformal time. }
The last term could be large $ \partial^i\partial^j\Phi \propto  {\delta \rho}/{\rho} \gg 1$ , where ${\delta \rho}/{\rho}$ is the perturbation in the matter density, However, we do not encounter such terms on the past light-cone, rather we have a projected mass density field, $\nabla^2_{\bot} \Phi,$  which is well-behaved due to the  weak gravitational lensing geometric term that regulate its amplitude.
Using equation \eqref{eq:positioneqn}, we calculate the perturbed position of the source
$
x^i(\bar{\eta},\bar{\x}) = \bar{x}^i(\bar{\eta}) +\delta x^i(\bar{\eta},\bar{\x}) \,,
$
where $\delta x^i$  denotes perturbation in the position, $\bar{\eta}$ and $\bar{\x}$ are the background conformal time and position respectively.  In the appendix \ref{deviationvector} we show how to re-map the background conformal time and position  in terms of the affine parameter $\lambda$ associated with the null geodesics 
\begin{eqnarray}\label{eq:photontracjectory}
x^i_{\rm{FLRW}} (\lambda_s,{\n})&\approx& \bar{d}_{A}\bigg\{ {n}^i +\frac{\delta x^i_{\bot}}{r_s} -\frac{1}{ r_s} \Delta x^j_{\bot} \nabla_{\bot j}\delta x^i_{\bot}
\\ \nonumber && 
+\bigg[ \frac{\delta x_{\p}}{r_s} - \frac{1}{r_s}\Delta x_{\p} \partial_{\p} \delta x_{\p}
-\frac{1}{r_s}\delta x_{\bot}^i \nabla_{\bot i}  \delta x_{\p}
%\qquad \qquad \qquad \qquad 
 - \left(1- \frac{1}{ r_s\HH_s}\right)
 \left( \delta \lambda -\Delta  x_{\p} \partial_{\p} \delta  \lambda   -\delta x^j_{\bot}\nabla_{\bot j} \delta \lambda \right){  \HH}\bigg]n^i
\bigg\}\,,
\end{eqnarray}
where  $r_s \equiv  (\lambda_o-\lambda_s)$ is the  comoving distance and $\bar{d}_A =a(\lambda_s)r_s$ is the area distance, $\HH$ is the conformal Hubble rate,  $\delta \lambda$ is the perturbation of the affine parameter and $ \Delta x_{\p}=\delta x_{\p} + \delta \lambda$.  Using equation \eqref{eq:photontracjectory}, we define the deviation vector within standard cosmology as 
\begin{eqnarray}
\frac{{{\xi}^i(\lambda_s,{\n})}}{\bar{d}_{A}} &\equiv& \frac{ {x^i_{\rm{FLRW}}(\lambda_s,{\n})-\bar{x}^i_{\rm{FLRW}}(\lambda_s)}}{\bar{d}_{A} }=  \frac{n^i \xi{\p}(\lambda_s,{\n}) + \xi_{\bot }^i(\lambda_s,{\n})}{\bar{d}_{A}}\,.
\end{eqnarray}
The orthogonal component or the deviation vector is given by
\begin{eqnarray}
%\frac{{\xi_{\bot }^{i}}\one (\lambda_s,{\n})}{\bar{d}_{A}}&=& \frac{\delta\one x^i_{\bot}}{r_s} \,, \label{eq:Delxone}\\
\frac{{\xi_{\bot }^{i}} (\lambda_s,{\n})}{\bar{d}_{A}} &=&\frac{\delta x^i_{\bot}}{r_s} -\frac{1}{ r_s} \Delta x^j_{\bot} \nabla_{\bot j}\delta  x^i_{\bot}\approx  2
\int^{r_s}_{0}\frac{(r_s-r)}{r_s}\nabla_{\bot}^i\Phi\one\d r\ \,.
\label{eq:Delxtwo}
\end{eqnarray}
Using this, we find that the gravitational lensing shear is given by {
\begin{eqnarray}
%\kappa\one(z_s,{\n})   & =&-\frac{1}{2}\int^{ r_s}_{0}\frac{( r- r_s) r}{ r_s}\nabla_{\bot }^2\left(\Psi\one+\Phi\one\right)\d r \,,\\
 \gamma\one_{ij}(z_s,{\n})&=&\int^{ r_s}_0\frac{({ r} -  r_s){ r}}{ r_s}\nabla_{\bot \<i}\nabla_{\bot j\>}(\Phi\one + \Psi\one) \d{ r}\,.
\end{eqnarray}
Note that in the low redshift limit, $\xi{\p}$ is the key term we focus on. {Therefore, in order to understand it much better,  we decompose it  further into geometric deformation  part $\delta r_{\rm{geo}}$ and  the radial or line of sight deformation part $\delta r_{\n}$:}
\begin{eqnarray}\label{eq:decomxiparallel}
\frac{{{\xi }{\p}} (\lambda_s,r{\n})}{\bar{d}_{A}} & \equiv & \frac{\delta r_{\rm{geo}}(\lambda_s,r{\n})}{r} +  \frac{\delta r_{\n}(\lambda_s,r{\n})}{r} \,,
\end{eqnarray}
where $\delta r_{\rm{geo}}$ is related to the Wrinkle surface effect described in \cite{Kaiser:2015iia}. {One can immediately isolate the terms that describe this effect from first three terms in the second line of equation \eqref{eq:photontracjectory}}
\begin{eqnarray}
% \frac{\delta\one r_{\rm{geo}}(\lambda_s,r{\n})}{r}&=&\frac{ \delta\one x_{\p}}{r_s}\,, \label{eq:deltar_geo1} %= - \frac{2}{r_s}\int^{r_s}_{0}\Phi\one\d r \,,
%\\
 \frac{\delta\ r_{\rm{geo}}(\lambda_s,r{\n})}{r}& =& \frac{\delta x_{\p}}{r_s}- \frac{1}{r_s}\Delta x_{\p} \partial_{\p} \delta x_{\p}-\frac{1}{r_s}\delta x_{\bot}^i \nabla_{\bot i} \delta x_{\p} \,. \label{eq:deltar_geo2}
\end{eqnarray}
The second term and third terms in equation \eqref{eq:deltar_geo2} are due to Post-Born correction~\cite{Ben-Dayan:2012lcv,Marozzi:2014kua,Bertacca:2014wga,DiDio:2014lka,Umeh:2020cag}.  Given the metric in equation \eqref{eq:metric}, $\delta r_{\rm{geo}}$ is given by
 \begin{eqnarray}
% \frac{\delta\one r_{\rm{geo}}}{r}&=& - \frac{2}{r_s}\int^{r_s}_{0}\Phi\one\d r \,,\label{eq:wrinklesurface1}\\
 \frac{\delta r_{\rm{geo}}}{r}& =&  -\frac{2}{r_s}  \int^{r_s}_{0}\left(\Phi + \Psi\right)\d r \label{eq:wrinklesurface2}
 \\ \nonumber  &&
 + \frac{4}{r_s} \int_{0}^{r_s} \d r' \int_{0}^{r'} \d r'' (r_s -r') 
 \nabla_{\bot i} \Phi (r'{\n}) \nabla_{\bot} ^i \Phi(r''{\n}) 
 + \frac{4}{r_s}\int^{r_s}_{0}{(r_s-r)}\nabla_{\bot}^i \Phi\int_{ 0}^{r_s}\frac{r}{r_s} \nabla_{\bot j }\Phi \d r \,.
\end{eqnarray}
 $\delta r_{\n}$ describes the deformation of the photon trajectory by the inhomogeneities along the line of sight. {One can pick out the terms that describe this effect from the last set of terms in the second line of  equation \eqref{eq:photontracjectory}}
\begin{eqnarray}\label{eq:radiallensing1}
 % \frac{\delta\one r_{\n}(\lambda_s,r{\n})}{r}  &=& -\left(1- \frac{1}{ r_s\HH_s}\right){  \HH_s} \delta\one \lambda\,,
 % \\
    \frac{\delta r_{\n}(\lambda_s,r{\n})}{r}  &=&- \left(1- \frac{1}{ r_s\HH_s}\right){\HH_s}\left( \delta \lambda -\Delta x_{\p} \partial_{\p} \delta \lambda  -\delta x^j_{\bot}\nabla_{\bot j} \delta \lambda \right)\,.
    \label{eq:radiallensing2}
\end{eqnarray}
  Earlier work by Ellis and  Solomons describes this contribution as radial lensing effect \cite{Ellis:1998qga}. {As we shall see in the next section that in redshift  space, it describes the parallax effect as earlier mentioned or more generally the Doppler lensing effect~\cite{Bacon:2014uja}.  This effect manifests as a displacement in the position of the  background source due to the fact that an observer is moving relative to a local over-density.  This effect has been used exclusively used to obtain the distance to the very nearby sources~\cite{Riess:2018byc}.} From its relationship with the tangential gravitational lensing convergence, $\kappa_{S}  =-\left({ \xi_{||}}/{r}-\kappa \right)$, it has opposite physical effect to $\kappa$~\cite{Bolejko:2012uj}.

%On the constant redshift  surface, it describes the Doppler lensing contribution~\cite{Bacon:2014uja}
\subsection{Local group Barycentric observer and affine parameter perturbations}\label{sec:redshiftperturbation}

Although the deviation vector is defined in terms  of the affine parameter which is monotonic when initialised at the observer, it is not an observable. What is observable is the  redshift of a source which  is defined as a ratio of the photon energy at the source, $E_s$, to the photon energy at the  observer location, $E_o$: 
\begin{eqnarray}
\left(1+z_{\rm{obs}}\right)
=\frac{E_s}{E_o} \,. 
\end{eqnarray}
{At late time,  gravitationally bound sources move relative to the cosmic rest frame. The peculiar motion contributes to the observed redshift of a source according to ~\cite{Davis:2010jq}
 \begin{eqnarray}\label{eq:obs-redshift}
 (1+ z_{\rm{obs}}) = (1+\bar{z})\left(1+ z_{\rm{pec}}^{\rm{o}}\right) \left(1+ z_{\rm{pec}}^{\rm{SN}}\right)  \,,
 %\left(1+ z_{\rm{\Phi}}^{\rm{helio}}\right) \left(1+ z_{\rm{\Phi}}^{\rm{SN}}\right)
 \end{eqnarray}
 where $\bar{z}$ is the redshift associated with the Hubble flow or the cosmological redshift(for the rest of the paper, we sometimes use $z$ to denote the cosmological redshifts especially in figures),     $z_{\rm{pec}}^{\rm{o}}$ is the  redshift associated with the motion of the heliocentric frame with respect to the cosmic rest frame and $z_{\rm{pec}}^{\rm{SN}}$ is the redshift associated with the peculiar motion of the source's frame with respect to the cosmic rest frame.
  In special relativity, $ z_{\rm{pec}}^{\rm{SN}}$ and $z_{\rm{pec}}^{\rm{helio}}$ are related to the peculiar velocity $v^X_i$  according to~\cite{Carr:2021lcj}
 \begin{eqnarray}
 (1+ z^{X}_{\rm{pec})} = \sqrt{\frac{1+ v^{X}_{\p}}{1-v^{X}_{\p}}}	\qquad {\rm{and }}\qquad (1+z^{X}_{\rm{pec}})=\frac{1}{\sqrt{1- {v^X_{\perp }}^{2}}}
 \end{eqnarray}
 where $v^{X}_i = n^i v_i^{X}$ is  line-of-sight direction component, $ {v^X_{\perp}}^{2} =  v^X_{\perp i} v^{X i}_{\perp }$   and   $v^X_{\perp i} = N_{i}{}^j v_j^{X}$ is the transverse component. 
 The relative velocity between the heliocentric frame and the cosmic rest frame receives contribution from several sources
 \begin{eqnarray}\label{eq:Vsun_CRF}
v_{\rm{sun-CRF}} =v_{\rm{sun-LSR}} + v_{\rm{LSR-GSR}}+{v_{\rm{GSR-LG}}  +  v_{\rm{LG-CRF}} }\,,
\end{eqnarray}
where $v_{\rm{sun-LSR}} $ is the motion of the sun relative to the nearby stars that define the dynamical Local Standard of Rest (LSR), $v_{\rm{LSR-GSR}}$ is the circular rotation of the LSR about the galactic center. $ v_{\rm{GSR-LG}}$ is the motion of the Galactic Standard of Rest (GSR)  or the  Galactic Rest Frame (GRF) relative to the Local Group (LG) centroid  and $v_{\rm{LG-CRF}} $ is the peculiar velocity of the LG with respect to the CRF.  Each of these relative velocity contributions defines a unique 4-velocity according to equation \eqref{eq:4velocity_def}. Each 4-velocity induces a unique coordinate system~\cite{1984ucp..book.....W}. Secondly, what we really want is $v_{\rm{sun-CRF}}$, we can't describe this within the cosmological perturbation theory on the expanding FLRW background (i.e equation \eqref{eq:metric}). We shall see in sub-section \ref{sec:tidaldeformation} that equation \eqref{eq:metric}) breaks down at about 1 Mpc.  Therefore, we  assume that the observer is located at the Barycenter of our local group and that the source is also positioned at the barycenter of its host halo or cluster. With these assumptions, we can re-write equation \eqref{eq:obs-redshift} as
  \begin{eqnarray}\label{eq:redshift1body}
(1+ z_{\rm{obs}})\approx (1+\bar{z})\left(1+ z_{\rm{pec}}^{\rm{LG}}\right) \left(1+ z_{\rm{pec}}^{\rm{SN,host}}\right)=\frac{E_s}{E_o} = {(- k_a u^a)_{s} \over (- k_b u^b)_{o}}\,,
\end{eqnarray}
where $z_{\rm{pec}}^{\rm{LG}}$ is the redshift associated with the motion of our local group and $z_{\rm{pec}}^{\rm{SN,host}}$ is the redshift associated with the host halo of the SNIa.  As we shall describe in detail in sub-section \ref{sec:allsky} that provided we correct for the monopole and dipole associated with the barycentric observers, we can always boost the final result back into the Heliocentric frame where the observer lives. 
% In order to obtain the corresponding expression for an observer in the Heliocentric frame, we simply need to boost to the Heliocentric frame.  
 We shall describe how this is done in detail elsewhere because this paper is already technically complicated, we do not intend to make it even more complicated.
 
Moreover, it is important to point out that the cosmological analysis of the SNIa samples in the Hubble flow are done as function of the cosmological redshift and not in terms of the heliocentric redshift~\cite{Peterson:2021hel}.   That is the cosmological analysis are not done in terms of the  observed redshift given in equation \eqref{eq:redshift1body}. Please see a more detailed discussion of this in \cite{Mohayaee:2021jzi}.  The constraint on  cosmological parameters would differ if the analysis is done in terms of $z_{\rm{obs}}$ especially when only a sub-sample of the low-s SNIa is used. Hence it is important to express all the SNIa on the constant redshift surface.
Details on how to obtain the cosmological redshifts from the observed redshift of sources for the Pantheon+ supernova samples were given in \cite{Carr:2021lcj}. The recent cosmological analysis of the Pantheon+ supernova samples made use of this technique~\cite{Brout:2022vxf}.
  
 On the theory side, this is easy to define. For example, to express our result in terms of the cosmological redshift or to define the surface of constant redshift, we  simply expand the full expression for the redshift in perturbation theory}
    \begin{eqnarray}\label{eq:phya}
 \frac{1}{(1+{z_{\rm{obs}}} )}&=& \frac{a({\lambda_{s}})}{a({\lambda_{o}})}\bigg[1+\left(-\HH \delta\one \lambda - \delta\one z\right)
 \\ \nonumber &&
 +\frac{1}{2} \left(-\HH  \delta\two \lambda- \delta\two z +2 (\delta\one z)^2 +2\HH  \delta\one z \delta\one \lambda 
 + \left(\frac{\d \HH  }{\d \lambda_{s}} + \HH^2 \right)(\delta\one \lambda)^2\right)+\mathcal{O}(\epsilon^3\bigg]\,.
 \end{eqnarray}
Imposing that the redshift is  entirely due to Hubble flow(cosmological redshift surface)  implies that 
 \begin{eqnarray}
 \delta\one \lambda &=&-\frac{\delta\one z}{  \HH}\,, \label{eq;consistencybody1}
 \\
  \delta\two\lambda&=&- \frac{1}{\HH}\left[\delta\two z 
  %-\frac{2}{\HH}\left[\Delta\one x_{\p}\partial_{\p} \delta\one z +2\delta\one x_{\bot}^i\nabla_{\bot i} \delta\one z\right]
  -(\delta\one z)^2 \left(1+ \frac{1}{\HH^2}\frac{\d\HH}{\d\lambda_{s}}\right)\right]\,.  \label{eq;consistencybody2}
 \end{eqnarray}
 This immediately fixed the perturbation in the affine parameter. {Plugging this back to equation \eqref{eq:radiallensing2}, we obtain the leading order terms contributing to the radial lensing  effect  at the constant redshift surface }
\begin{eqnarray}\label{eq:Dopplerlensing}
 %\frac{\delta\one r_{\n}}{r}
% &=& -\left(1 - \frac{1}{r_{s} \HH_{s}}\right)\left(\partial_{\|}{v\one_s}-  \partial_{\|}v\one_o\right)\,, \label{eq:radial_lensing1}\\ 
  \frac{\delta\ r_{\n}}{r}
  &\approx&-\left(1 - \frac{1}{r_{s} \HH_{s}}\right)  \bigg[\left(\partial_{\|}{v_s}-  \partial_{\|}v_o\right)
+\frac{1}{\HH_{s}} \left(\partial_{\|}{v_s}-  \partial_{\|}v_o\right) \partial^2\p{v_s}
%\\  \nonumber&&
+2\nabla_{\bot j}\partial_{\|}{v_s}\int^{r_s}_0 {(r - r_s)}\nabla_{\bot}^i\Phi \d r \bigg]\,. \label{eq:radial_lensing2} 
\end{eqnarray}
where we have focused on the leading order terms. 
{The leading order approximation is determined by considering the number of spatial derivatives~\cite{Umeh:2016nuh}.
For example, at  the linear order in perturbation theory,  terms with two spatial derivatives(it could be angular, or radial(derivative along the line of sight)) of the gravitational potential or the velocity potential are dominant terms. These  are terms which are traditionally classified as Newtonian terms.  For example, matter density,  Kaiser redshift space distortion term.
Terms with one or zero spatial derivatives are sub-dominant and are classified as general relativistic  terms. Under this scheme, the first term in equation \eqref{eq:coeff2} is classified as general relativistic term. 
Similarly, at the second-order in perturbation theory,  the same scheme applies to the intrinsic second-order term.  For the terms quadratic in the first-order terms, the Newtonian terms contain four spatial derivatives in total while the general relativistic term terms are those with less than four spatial derivatives. For example in equation \eqref{eq:coeff2}, the leading order terms at second order are those with three partial derivatives of the gravitational potential or velocity potential along the line of sight or angular derivatives. This is the reason why we sometimes describe the terms in equation \eqref{eq:coeff2}  as general relativistic effects. }

Comparing equation \eqref{eq:radial_lensing2}  to equation \eqref{eq:wrinklesurface2}, we observe that the geometric deformation terms are sub-dominant at low redshift, hence, the  leading order contribution comes from  the radial lensing effect or the parallax effect.
Finally the leading order approximation we will be studying in the subsequent sub-section is
 \begin{eqnarray}\label{eq:coeff}
\frac{{{\xi }{\p}(\bar{z},{\n})}}{\bar{d}_A} & = & \frac{\delta r_{\rm{geo}}}{r} +  \frac{\delta r_{\n}}{r} \approx\frac{\delta r_{\n}}{r} 
\\
&\approx&
-\left(1 - \frac{1}{r_{s} \HH_{s}}\right)  \bigg[\left(\partial_{\|}{v_s}-  \partial_{\|}v_o\right)
+\frac{1}{\HH_{s}} \left(\partial_{\|}{v_s}-  \partial_{\|}v_o\right) \partial^2\p{v_s}
%\\  \nonumber&&
+2\nabla_{\bot j}\partial_{\|}{v_s}\int^{r_s}_0 {(r - r_s)}\nabla_{\bot}^i\Phi \d r \bigg]\,.
\label{eq:coeff2}
 \end{eqnarray}
 {The second and the third terms of equation \eqref{eq:coeff2} are what we study in detail in sub-section \ref{sec:allsky}.}

%%%%%%%%%%%%%%%%%%%%%%%%%%%%%%%%%%%%%%%%%%%%%%%%%%%%%%%%%%%%%%%%%x

%\subsection{Monopole of the area distance at constant redshift surface}
 \subsection{Monopole of the area distance and full sky spherical harmonics decomposition}\label{sec:allsky}

Our interest is to calculate the monopole or the all sky average or direction average  of the area distance
 \begin{eqnarray} 
 \averageA{  {d}_A(\bar{z},{\n})}&=&\bar{d}_A(\bar{z})  \bigg[1+\averageA{ \frac{ \xi{\p}}{\bar{d}_A } }+ \averageA{\kappa} 
 -\frac{1}{4}\bigg(\averageA{\gamma_{ij} \gamma^{ij}}-\averageA{\omega_{ij}\omega^{ij} }\bigg) \bigg]\,.
 \label{eq:allskyaverage}
 \end{eqnarray}
 where $ \averageA{\cdots}$ denotes the monopole on a constant redshift ($z$) sphere
 \begin{eqnarray}
\averageA{X(\bar{z})}&\equiv&% \frac{1}{4\pi} \int \d^2 \theta X(\bar{z},\theta) =
 \int \d {\n} X(\bar{z},{\n})\,.
\end{eqnarray}
The  all sky average of the lensing convergence $ \averageA{\kappa}$  vanishes because $\kappa$ is a total divergence  of a 2-vector on the screen space 
$\averageA{\kappa} =- \int \d \Omega\, \nabla_{\bot }^B \xi_{\bot  B}/2 = 0 \,.$
The twist vanishes for purely scalar perturbation at leading order~\cite{Marozzi:2016qxl}, hence $\averageA{\omega_{ij}{\omega}^{ij}}=0$ leaving
\begin{eqnarray}\label{eq:averageDA}
\averageA{  {d}_A(\bar{z},{\n})}&=&\bar{d}_A(\bar{z})\bigg[1+  \averageA{\frac{\xi{\p}}{\bar{d}_A} }
-\frac{1}{4}\left(\averageA{ \gamma_{ij} {\gamma}^{ij}} \right)\bigg]\,.
\end{eqnarray}
Note that $\averageA{ { \xi{\p}}/{\bar{d}_A} }\neq 0$, it is a general relativistic effect. We will show that in the low redshift limit $\averageA{ { \xi{\p}}/{\bar{d}_A} }$ dominates: $\averageA{  {d}_A(\bar{z},{\n})}\simeq\bar{d}_A(\bar{z})\big[1+  \averageA{{\xi{\p}}/{\bar{d}_A} }\big]$ and its contribution is non-negative. At high redshift,  equation \eqref{eq:allskyaverage} reduces to 
 \begin{eqnarray}\label{eq:Areadisatnce6}
\averageA{{d}_{A}(\bar{z},{\n})}&\simeq & \bar{d}_A(\bar{z})\bigg[1-\frac{1}{4}
\averageA{ \gamma_{ij} {\gamma}^{ij}}\bigg]  
\simeq  \bar{d}_A(\bar{z})\bigg[1-\frac{1}{2}\averageA{ \left(\kappa\right)^2}\bigg] \,,
\end{eqnarray}
where $\averageA{\kappa^2}\approx\averageA{\gamma_{ij} \gamma^{ij}} /2$ on small scales.   
This is a well-known result~\cite{Clarkson:2014pda,Kaiser:2015iia}. 
% The average  of equation \eqref{eq:Areadisatnce6} was discussed in \cite{Clarkson:2014pda,Kaiser:2015iia,Breton:2020puw}, however, we focus on the direction average here,  the effects we discuss will apply in that case as well. 

%In the second equality,  we made the replacement $(\gamma_{ij} \sim \kappa N_{ij}$).  %In general, the all-sky average of area distance is given by
{In order to evaluate the all sky average of the perturbation to the area distance given in equation \eqref{eq:averageDA}, we split equation \eqref{eq:coeff2} (i.e $\averageA{ { \xi{\p}}/{\bar{d}_A } }$)  further into two parts in order to  improve  clarity of presentation }
\begin{eqnarray}\label{eq:averagedeflection}
  \averageA{\frac{\xi{\p}}{\bar{d}_A} } \equiv   \averageA{\frac{\xi{\p}}{\bar{d}_A} }^{A} +  \averageA{\frac{\xi{\p}}{\bar{d}_A} }^{B}\,,
\end{eqnarray} 
where $ \averageA{{\xi{\p}}/{\bar{d}_A} }^{A}$ and $\averageA{\xi{\p}/{\bar{d}_A} }^{B}$  are the monopole of the second and third terms in equation \eqref{eq:coeff2}  respectively.  
%The first term is zero on average. 
%These are the post-Born radial distortion terms due to LoS peculiar velocity  and gravitational deflection respectively.
%In the plane parallel limit, $ \averageA{{\xi{\p}}/{\bar{d}_A} }^{A}$ and  $\averageA{\xi{\p}/{\bar{d}_A} }^{B}$ vanish if the terms evaluated at the observer  position are set to zero a priori(please see Appendix \ref{sec:harmonicsdecomp} for details). Earlier studies have missed this effect simply because they neglected terms evaluated at the observer position and made a plane parallel approximation~\cite{BenDayan:2012pp,Ben-Dayan:2012uam,Fleury:2016fda}.  The importance of the terms evaluated at the observer position in regulating the  infra-red divergences in the luminosity distance was discussed here~\cite{Biern:2016kys}.  Its implication on the two-point correlation was discussed in \cite{Grimm:2020ays}.
%Expanding equations \eqref{eq:averagedeflection} and $\averageA{\gamma_{ij} { \gamma}^{ij}}$  in  full-sky spherical harmonics  we find 
Now let's go through the full sky spherical harmonic expansion of each term in detail.   Firstly, we separate the terms contributing to $ \averageA{{\xi{\p}}/{\bar{d}_A} }^{A}$  into terms evaluated at the same source position from the term evaluated at both source and observer positions:
{
 \begin{eqnarray}
   \averageA{\frac{\xi{\p}}{\bar{d}_A} }^{A}%&\approx& - \left(1 - \frac{1}{ r_{s} \HH_{s}}\right) \averageA{
%\frac{1}{\HH_{s}} \left(\partial_{\|}{v\one_s}-  \partial_{\|}v\one_o\right) \partial^2\p{v\one_s}}\,,\\
&\approx&  - \left(1 - \frac{1}{ r_{s} \HH_{s}}\right) \frac{1}{\HH_{s}}\bigg[\averageA{
 \partial_{\|}{v\one_s} \partial^2\p{v\one_s}} -\averageA{
 \partial_{\|}{v\one_o} \partial^2\p{v\one_s}} \bigg] \,.\label{eq:AveragexiA}
 \end{eqnarray}
The all sky average of the term evaluated at the same source position i.e  $\averageA{
 \partial_{\|}{v\one_s} \partial^2\p{v\one_s}} $ is zero by symmetry. This implies that all the contribution to the monopole of the  area distance from $\averageA{{\xi{\p}}/{\bar{d}_A} }^{A}$ is due to $\averageA{
 \partial_{\|}{v\one_o} \partial^2\p{v\one_s}}$:
\begin{eqnarray}\label{eq:parallax_effect3}
 \averageA{ \frac{\xi{\p}}{d_{A}} }^{A} &=& \left(1 - \frac{1}{r_{s} \HH_{s}}\right)
     \averageA{ \partial_{\|}v_o \partial^2\p{v_s}}\,.
\end{eqnarray}
Note that this contribution is dependent on the observer peculiar velocity and becomes important at very low redshifts. That's why we refer to as  parallax effect. It can be used to determine not just Stellar distances and distances to nearby extra-galactic sources~\cite{Paine:2019vep}.
Expanding equation \eqref{eq:parallax_effect3} in full sky spherical harmonics leads to 
\begin{eqnarray}
 \averageA{ \frac{\xi{\p}}{d_{A}} }^{A} &=& \left(1 - \frac{1}{ r_{s} \HH_{s}}\right)  (D(z)\HH(z)f(z))(D(z_o)\HH(z_o) f(z_o)) \sum_{\ell = 0}^{\ell_{\rm{max}}} (2\ell+1)\int \frac{\d k}{2\pi^2} k P_{m}(k)  j'_{\ell}(kr_o) j''_{\ell}(kr)\,.
 \label{eq:shericalharmonicsxiobs}
\end{eqnarray}
Please see Appendix \ref{sec:harmonicsdecomp} for details on how to perform the spherical harmonics decomposition. 
Ordinarily, equation \eqref{eq:shericalharmonicsxiobs} would have been the final result we are looking for but we have to remember that due to the restriction(barycentric observer approximation) or the cosmological perturbation theory constraint, i.e we positioned the observer at the barycenter of our local group. The $\ell =0$ and $\ell =1$ modes carry this information. Therefore, when we sum from $\ell =0$ to $\ell =\infty$, we are calculating the monopole of the area distance that the local group barycentric observer will measure.  In order to remove this obvious coordinate dependence of the final result, we adopt the strategy used in the analysis of the CMB temperature anisotropies~\cite{Hu:1997hp}.

The strategy adopted in the treatment of the CMB temperature anisotropies in the presence of  large peculiar velocity contribution to the monopole and the dipole is to expand all the terms(i.e equation \eqref{eq:AveragexiA}) in full sky spherical harmonics and them sum from $\ell =2$ to $\ell =\infty$~\cite{Hu:1997hp,Planck:2020qil}. Using the correspondence between the spherical harmonics and symmetric trace-free tensor, one can show that $\ell=0$ and $\ell =1$ modes are coordinate dependent \cite{Pirani2,Planck:2020qil}.
% In the second equality, we dropped the peculiar velocity evaluated at the observer position. When expanded in spherical harmonics, it can only contribute at $\ell=1$. Dropping it at this point implies that we can only sum the multipoles from $\ell =2$.
% Now, we describe how we perform the full-sky spherical harmonics of decomposition of equation \eqref{eq:AveragexiA}. 
%
We showed in Appendix \ref{sec:harmonicsdecomp} that the following correspondence holds at low redshift
\begin{eqnarray}\label{eq:correspond}
\overline{\averageA{\partial_{||} v\one_{s} \partial^2_{||} v\one_{s}} } -\overline{ \averageA{\partial_{\|}{v\one_o} \partial^2\p{v\one_s}}}&\approx&\overline{\averageA{\partial_{||} v\one_{s} \partial^2_{||} v\one_{s}} } \,, \qquad{\rm{with}}\qquad \ell_{\rm{min}} = 2.
\\
 &=&(D(z)\HH(z)f(z))^2 \sum_{\ell = 2}^{\ell_{\rm{max}}} (2\ell+1)\int \frac{\d k}{2\pi^2} k P_{m}(k)  j'_{\ell}(kr) j''_{\ell}(kr) \,.
%&=& { (D(z)\HH(z)f(z))(D(z_o)\HH(z_o) f(z_o)) \sum_{\ell = 0}^{\ell_{\rm{max}}} (2\ell+1)\int \frac{\d k}{2\pi^2} k P_{m}(k)  j'_{\ell}(kr_o) j''_{\ell}(kr)}\,.
%\qquad
\end{eqnarray}
We made use of this correspondence for computational efficiency.
}
Using this strategy, we find that the full-sky spherical harmonics of decomposition of the rest of terms are
\begin{eqnarray}
\averageA{ \frac{\xi{\p}}{d_{A}} }^{A} &=&- \left(1 - \frac{1}{ r_{s} \HH_{s}}\right)  \HH(z_s)( f(z_s)D(z_s))^2 \sum_{\ell =2}^{\ell_{\rm{max}}} (2\ell+1)\int \frac{d k_1}{2\pi^2} k_{1}P_{m}(k_1) j'_{\ell}(k_1  r_s)j''_{\ell}(k_1  r_s)\,,
\label{eq:xiA2}
\\
 \averageA{ \frac{\xi{\p}}{d_{A}} }^{B} &=& -\left(1 - \frac{1}{ r_{s} \HH_{s}}\right)\sum_{\ell =1}^{\ell_{\rm{max}}}  (2\ell+1)     \frac{(\ell +1)!}{(\ell-1)!}  {\HH(z_s)}{f(z_s) D(z_s)} 
   %\\ &&\times
\left[\frac{1}{\pi}\right]^2 
   \int\frac{d k_1}{ k_{1}}P_{m}(k_1) j'_{\ell}(k_1 r_s) \mathcal{I}_{\ell}(k_1, r_s)
\,,
\label{eq:xiB2}
\\  
\averageA{\gamma_{ij} { \gamma}^{ij}}&=&  \left[\frac{2}{\pi^2}\right] \sum_{\ell =2}^{\ell_{\rm{max}}}  (2\ell+1)   \frac{(\ell +2)!}{(\ell-2)!} 
\int  \frac{d k_1}{k_1^2} P_{m}(k_1)\mathcal{I}_{\ell}(k_1, r_s)\mathcal{I}_{\ell}(k_1, r_s)\,,
\label{eq:gammasq}
\end{eqnarray}
where  $D$ is the growth factor for the density perturbation and $f$ is the growth rate, $j_{\ell}(x)$ is the spherical Bessel function and $j'_{\ell}(x)$  and   $j''_{\ell}(x)$ are first and second  derivatives of the spherical Bessel  function  of order $\ell$ . $P_{m}$ is the matter power spectrum,  we compute $P_m$ using the CAMB code~\cite{Lewis:1999bs},  $\mathcal{I}_{\ell}$ is the integral along the line of sight
\begin{eqnarray}
\mathcal{I}_{\ell}(k, r_s) =  -\frac{3}{2}\Omega_{m0}H_0^2 \int_{0}^{ r_s}{\d} r \bigg[ \frac{( r- r_s)}{ r_s r}(1+z) D(z) j_{\ell} (k_1 r) \bigg]\,.
\end{eqnarray}
Again equations \eqref{eq:xiA2} and  \eqref{eq:xiB2} vanish in the plane-parallel limit when the observer dependent term is neglected. So it is important that a  full-sky spherical harmonic decomposition is employed.  
%We made use of the Total Angular Momentum approach (TAM) introduced in the analysis of the CMB to calculate their contribution to the monopole of the area distance~\cite{Hu:1997hp}. 
%
%
  \begin{figure}[h]
\centering % \begin{center}/\end{center} takes some additional vertical spaceMultipolesxiAwith.png
\includegraphics[width=80mm,height=40mm] {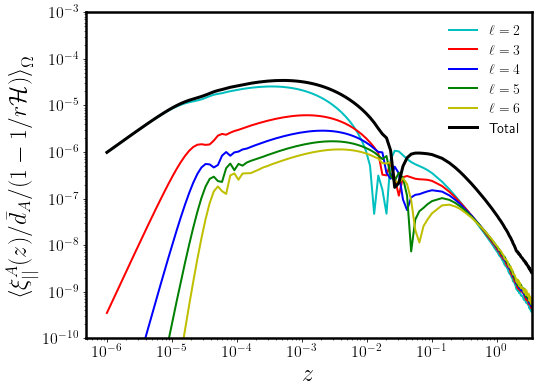}
\includegraphics[width=80mm,height=40mm] {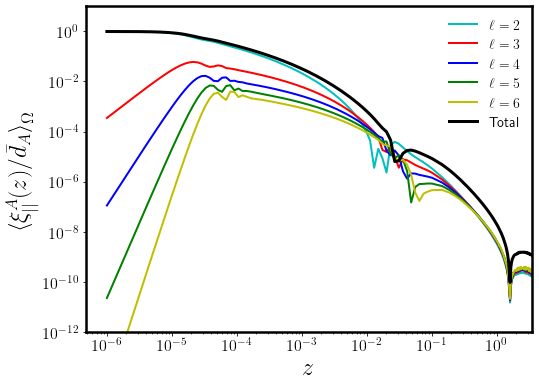}
\includegraphics[width=80mm,height=40mm] {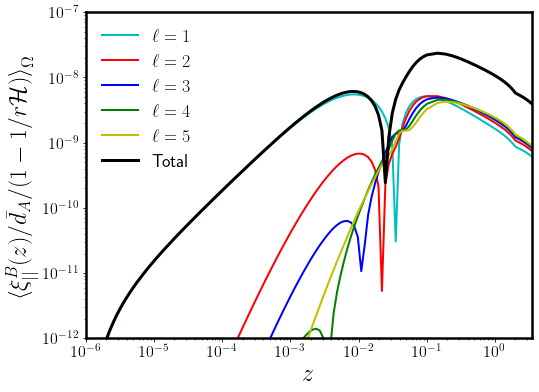}
\includegraphics[width=80mm,height=40mm] {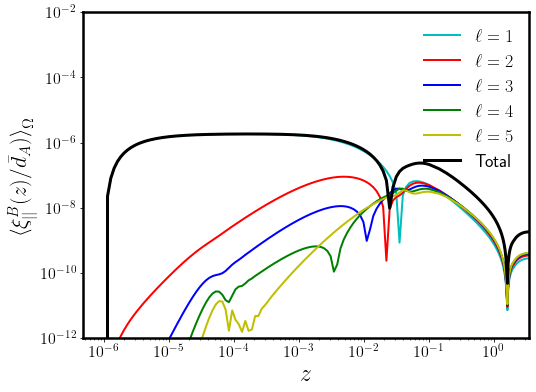}
\caption{\label{fig:BiasEv} The top right and left panels show the monopole of the Post-Born radial distortion due to LoS peculiar velocity, ( i.e. $ \averageA{{\xi{\p}}/{\bar{d}_A} }^{A}$) with and without the geometric boost factor, i.e $\left(1 - {1}/{ r(z_s) \HH(z_s)}\right)$ respectively.
Similarly, the bottom right and left panels show the monopole of the Post-Born radial distortion due to
 gravitational deflection ($\averageA{\xi{\p}/{\bar{d}_A} }^{B}$).
 The dark curve denotes the sum of the first few $\ell$s. On the left panels, the curves appear to change signs around $z =0.001$ and $ z= 0.01$. We find this feature disappears as we increase $\ell_{\rm{max}}$. }
\end{figure}
%The impact of weak gravitational lensing shear  to the monopole becomes important on small scales \cite{Montanari:2015rga} or at large $\ell$ and high redshift.
At large $\ell$,   the three integrals in $\averageA{\gamma\one_{ij} { \gamma\one}^{ij}}$ maybe reduced to one using the Limber approximation \cite{Limber:1954zz}:
\begin{eqnarray}\label{eq:Limberapprox}
\frac{2}{\pi} \int k^2 {\d} k f(k) j_{\ell}(k r) j_{\ell}(k r') = \frac{\delta( r -  r')}{ r^2} f\left(\frac{\ell +1/2)}{ r}\right)\,.
\end{eqnarray}
Implementing these is equation \eqref{eq:gammasq} and performing the delta function integration leads to 
\begin{eqnarray}\label{eq:Limberexp}
\averageA{ \gamma\one_{ij} { \gamma\one}^{ij}}&\approx &\frac{9}{4}\frac{\Omega^2_{m0}}{\pi}\sum_{\ell =2}^{\ell_{\rm{max}}}  (2\ell+1)  \frac{(\ell_1+2)!}{(\ell_1-2)!} 
\left(\frac{H_0}{(\ell+ 1/2)}\right)^4
%\\ \nonumber &&
\int_0^{ r_s}{ \d}  r  {\left[\frac{D(z)(1+z) ( r- r_s)}{ r_s}\right]^2}P_{m}\left(k\right)\,,
\end{eqnarray}
where the momentum dependence is substituted in terms of $ r$ using $k = (\ell +1/2)/ r$.  This is valid provided $\ell >10$ \cite{Bernardeau:2011tc}.
  \begin{figure}[h]
\centering % \begin{center}/\end{center} takes some additional vertical space
\includegraphics[width=75mm,height=60mm] {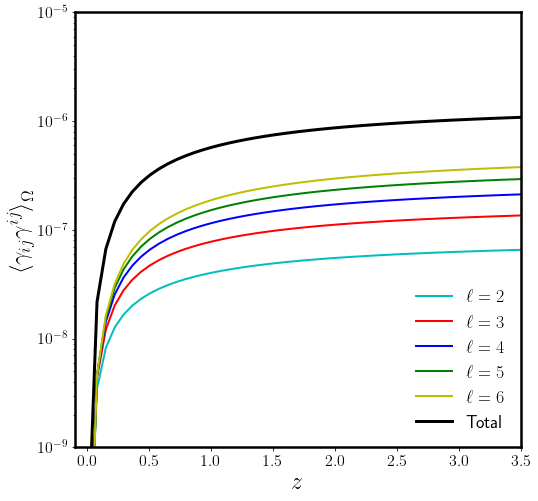}
\includegraphics[width=75mm,height=60mm] {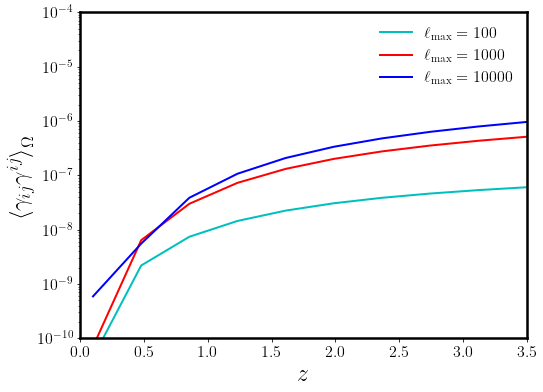}
\caption{\label{fig:leningshear}  On the left panel, we show the first few large scale modes contributing to the monopole of the weak gravitational lensing shear contribution to the area distance. The sum of the first few $\ell$s is denoted with a dark curve. On the right panel, we show the result of using the Limber approximation (equation \eqref{eq:Limberexp}) to calculate the monopole of the weak gravitational lensing shear contribution. For a more accurate result at high redshift, one needs to sum up to $\ell_{\rm{max}} = 10^6$~\cite{Clarkson:2014pda}. We did not bother to do this since our focus is on very small redshifts.   }
\end{figure}
The  contribution of each of the terms  given in equations \eqref{eq:xiA2} and \eqref{eq:xiB2}  are shown in figure \ref{fig:BiasEv}. We study how the coefficient $\left(1 - {1}/{ r(z_s) \HH(z_s)}\right)$ boosts the contribution in the limit of zero redshift. In the limit $z\to 0$, $ \averageA{ {\xi{\p}}/{d_{A}} }^{A} \to 1 $ indicating that the strong gravitational lensing effect must be included, hence, the single parameter perturbation theory discussion we have adopted becomes unreliable.  We provide steps on what this means and how to handle it in sub-section \ref{sec:tidaldeformation}.

We show the contribution of the weak gravitational lensing shear in figure \ref{fig:leningshear}. On the left panel, we show the contribution of the weak gravitational lensing shear from the first few multipole moments. We made use of equation \eqref{eq:gammasq} to calculate it, i.e without making the Limber approximation. While on the right panel, we use the Limber approximation (equation \eqref{eq:Limberexp}) to calculate a few representative multipole moments. 
%Note that the results shown in figure \ref{fig:leningshear} are in agreement with Weinberg’s proof that the apparent magnitude of a distant object is unaffected by gravitational lensing~\cite{1976ApJ...208L...1W}.  
One key point to note from this is that the impact of cosmic structures may be neglected at high redshift and area distance based on the background FLRW could be used without worrying about any bias from the tangential lensing terms. 
  \begin{figure}[h]
\centering 
\includegraphics[width=100mm,height=70mm] {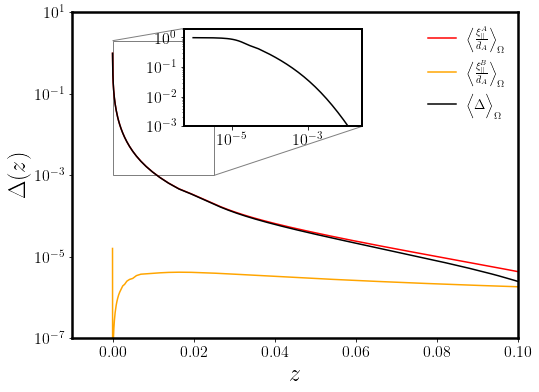}
\caption{\label{fig:totalDA}  
We show the monopole of the fluctuation in the area distance ${\Delta(z)} $ as a function of redshift for the key term that dominates at small redshift. The inset is a log-log plot of the monopole focussing on the $z\le 0.023$ redshift range. 
For this plot we set $\ell_{\rm{max}} = 100$.
}
\end{figure}

We show in figure \ref{fig:totalDA} the fractional difference between the monopole of the area distance and the FLRW background limit: ${\Delta(z)} = \left(\averageA{d_{A}(z,{\n}) }- \bar{d}_{A}(z)\right)/{\bar{d}_{A}(z)}$. At about $z =0.023$, we find about $1\%$ correction to the area distance from large scale structures. Although, we considered only the leading order terms in the limit $z\to0$, we do not think that including the other terms we neglected will change significantly this conclusion. Thus, this shows that using the background FLRW space time to interpret the SNIa sample is valid. This is in agreement with related studies in this direction~\cite{Ben-Dayan:2014swa,Fleury:2016fda}.
However, as $z\to0$, i.e the regime where $ \averageA{ {\xi{\p}}/{d_{A}} }^{A} $, dominates. We see from the top-left panel of figure \ref{fig:BiasEv} and also in figure \ref{fig:totalDA} that the correction increases dramatically to about $100\%$.
% It plateaus at about the redshift $z=5\times10^{-5}$. This is likely an indication of a break down of cosmological perturbation theory treatment. Or that sources with redshift less than some $z_{\rm{\star}}=5\times 10^{-5}$ are not in the Hubble flow.  
The part of the contribution to the radial lensing due to the gravitational deflection is sub-dominant and may be neglected in this limit as well. We checked that going to very high $\ell$ does not change this conclusion.
This makes including the contribution of $ \averageA{{\xi{\p}}/{d_{A}} }^{A}$ to the area distance in the distance likelihood code much simpler \cite{Greene:2021shv}. Note that the computation of $\averageA{\xi_{\|}/\bar{d}_{A}}^A$ involves only one momentum integral which can easily be optimised. Our crude implementation in Python takes less than a second to run.

\subsection{Imprints of tidal deformations on the area distance at low redshift} \label{sec:tidaldeformation}

On the right panel of figure \ref{fig:leningshear}, we show how the coefficient, i.e  $\left(1 - {1}/{ r(z_s) \HH(z_s)}\right)$ boosts the contribution of the post-Born terms in the limit $z\to0$.
On its own,  this term diverges  in this limit 
 \begin{eqnarray}
\left(1 - \frac{1}{ r(z_s) \HH(z_s)}\right)\big|_{z_s \rightarrow 0 } \to- \frac{1}{ r(z_s) \HH(z_s)} \to -\infty\,.
\end{eqnarray}
The contribution of $\left( \partial_{\|}{v\one_s}\partial^2\p{v\one_s}/{\HH}\right)$ (without the coefficient) to the monopole of the area distance is less than $0.1\%$.   It is this coefficient  that boosts the contribution  to almost 100\% in the limit $z \to 0.0$.
{So what is going on?} Let's examine the origin of this term first. From equation \eqref{eq:averagedeflection}, in the limit $z\to0$, the radial distortion is given by
\begin{eqnarray}\label{eq:dominant}
 \lim_{z\to0} \averageA{\frac{\xi{\p}}{\bar{d}_A} }\simeq  \averageA{\frac{\xi{\p}}{\bar{d}_A} }^{A}&\approx&   \frac{1}{ r} \averageA{
\frac{1}{\HH^2} \left(\partial_{\|}{v_s}-  \partial_{\|}v_o\right)\partial^2\p{v\one}}\,.
\end{eqnarray}
The most important thing about this term is that it nonzero because the null geodesics converge at the observer position; $-\nabla_{\bot i} k^i_{\bot} = \partial_{\bot i} n^{i}_{\bot} = {2}/{r}$. This is the  where the coefficient $1/r$ which boosts the amplitude of the term contained in the angle brackets comes from. %It is responsible for the large correction shown  in figure \ref{fig:totalDA}.
It determines the effective amplitude of the correction  to the  background  area distance.  
Taking the zero redshift limit of equation \eqref{eq:dominant} gives 
\begin{eqnarray}\label{eq:harmonic-decomp}
\lim_{z\to0}\averageA{\frac{\xi_{||}}{\bar{d}_{A}}}  \approx \lim_{z_s\to0} \frac{5f^2(z_s)D^2(z_s)}{r(z_s) }\sum_{\ell = 2}^{\ell_{\rm{max}}}(2\ell+1) \int  \frac{d k_1}{2\pi^2} \left[k_{1}P_{m}(k_1)
 j'_{\ell}(k_1 r_s)j''_{\ell}(k_1 r_s)\right]\to \frac{0}{0}\,.
\end{eqnarray}
The effective contribution from $\averageA{ {\xi{\p}}/{\bar{d}_{A}} }$ is indeterminate. However, applying L'Hopital's rule, we find  that $\averageA{{\xi_{||}}/{\bar{d}_{A}}}$ is conditionally finite
 \begin{eqnarray}\label{eq:differentiable}
 \lim_{z\to 0}  \averageA{ \frac{\xi{\p}}{\bar{d}_{A}} }=5 f^2(0) 
 \int\frac{d k_1}{2\pi^2} k^2_{1}P_{m}(k_1)
 j''_{2}(0)j''_{2}(0)\ \approx \frac{4}{45} f^2(0)  \sigma^2_R\,,
 \end{eqnarray}
 where we made use  of $j''_{2}( 0) = {2}/{15}$.  $ \sigma^2_R$ is the variance in dark matter density field
\begin{eqnarray}\label{eq:cosmic_variance}
 \sigma^2_R&=& \frac{1}{2\pi^2} \int_{0}^{k_{\rm{UV}}} {\d} k \,k^2 P_{\rm{m}}(k)  
=\frac{1}{2\pi^2} \int_{0}^{\infty} {\d} k \,(kW_{\rm{Th}}(k R))^2 P_{\rm{m}}(k) \,.
\end{eqnarray}
In the second equality, we have introduced a top-hat  window function, $W_{\rm{Th}}$, in  lieu of a UV dependent momentum integral. 
This means that  $\averageA{{\xi{\p}}/{\bar{d}_{A}} } $ exists in the limit $z\to0$,  provided that  the smoothing scale, $R$, is non-zero. %The key issue to be addressed is the question of the most physically justified  scale  to fix $R$ at. We discuss this in detail in section \ref{sec:causalhorizon}. 

It is more informative to express equation \eqref{eq:differentiable} in the following form
\begin{eqnarray}\label{eq:averagetidal}
 \lim_{z\to 0}  \averageA{ \frac{\xi{\p}}{\bar{d}_{A}} } \approx \frac{4}{45}f^2(0)\sigma^2_{R}= \frac{2}{15}\averageA{\frac{ \sigma_{ij}\sigma^{ij}}{H^2_0}}\,.
\end{eqnarray}
In the last equality, we have introduced the shear tensor, $\sigma_{ij}$. It is related to the peculiar velocity $v^i$ according
$\sigma_{ij} = \partial_{i}  v_{ j } -\delta^{ij}\partial_i v_{ j }/3.$
$\sigma_{ij}$ is symmetric and trace-free. 
It describes the cumulative effect of the tidal field(environment) from the initial seed  time to today \cite{Bertschinger:1994nc}
 \begin{eqnarray}
{\sigma}_{ij}(\eta_{0}, {\x}) &\sim&  -  \int_{\tau_{\rm{ini}}}^{\eta_{0}}  {d\eta'}{E}_{ij}(\eta', {\x}')  \,,%=-  \int_{\tau_{\rm{ini}}}^{\eta_{0}}  {d\eta'} \partial_{\<i}\partial_{j\>} \Phi(\eta', {\x})\,,
\end{eqnarray}
where $E_{ij}$  is  the electric part of the Weyl tensor.
%We made use of perturbed metric given equation \eqref{eq:metric} in the second equality. 
$E_{ij}$ describes how nearby geodesics tear apart from each other~\cite{Ellis:1990gi,Hahn:2006mk,Libeskind:2017tun}. 
Essentially, ${\sigma}_{ij}$ carries information about cumulative tidal deformation experienced by our local group due to its gravitational interaction within itself and its nearest neighbours. 

%\red{Therefore, equation \eqref{eq:differentiable} describes the effective tidal deformation experienced within a domain of radius $R$ around the observer. The key question to be addressed shortly is the question about the most physically justified scale to fix $R$ at.}
One key concern about the zero-redshift limit in equation \eqref{eq:averagetidal} or the physical significance of the very low redshift features in figure \ref{fig:totalDA} is that redshift becomes multi-valued when a light beam passes through a collapsing region. Some directions could be expanding while some will be collapsing \cite{2012MNRAS.425.2049H,Dalal:2020mjw}. This kind of situation is usually associated with the presence of a focal point, i.e a point where the area of the beam cross-section goes through zero~\cite{Perlick:2010zh}. In the context of gravitational lensing, it is known as caustics~\cite{Ellis:1998ha,Ellis:1998qga}. From the differential geometry point of view, it is an indication of a breakdown of the global coordinates, i.e a bundle of light rays can no longer be described a single family of null geodesics beyond a focal point~\cite{Witten:2019qhl}. This implies that below the focal point, the distance-cosmological redshift relation breaks down or cannot be trusted, especially the mapping in equations \eqref{eq;consistencybody1} and \eqref{eq;consistencybody2}.

Using the halo model, we showed in \cite{Umeh:2022prn} that the physical boundary of halos(i.e the splashback radius \cite{Adhikari:2014lna,Zuercher:2018prq,Murata:2020enz}) are usually located closer to the observer than the zero-velocity surface. Stated differently,  the splashback radius of the host halo is always less than the radius of the zero-velocity surface when measured with respect to an observer at $r=0$(see the right panel of figure \ref{fig:Error}).
We define the radius of the zero-velocity surface as the comoving distance where the divergence of the 4-velocity vector of the observer geodesic vanish
\begin{eqnarray}\label{eq:horizon}
R \ge - \frac{c}{3 H_0}  \frac{\d \ln \rho}{\d \ln r}\,,
\end{eqnarray}
where $c$ is the speed of light and $\rho$ is the local matter density. It is possible to calculate equation \eqref{eq:horizon} given some halo profiles~\cite{1989AandA...223...89E,Navarro:1995iw} with the outer profile given by the mean matter density~\cite{Diemer:2017bwl}.  But there are observational constraints for our local group which falls within the following range $R\sim(0.95- 1.05) {\rm{Mpc}} $~\cite{1999A&ARv...9..273V,Li:2007eg,Karachentsev:2008st}. Therefore, the smoothing scale in equation \eqref{eq:averagetidal} must not be less than $R$. The features in figure \ref{fig:totalDA} makes sense only when the distance between the observer and the source is greater than $R$. Using the background FLRW distance redshift relation, this corresponds to about $z_{\rm{cut}} = 2.4\times 10^{-4}$.

 {Finally, one of the key results we report here is that the area distance based on the background FLRW spacetime is insufficient to describe distances in a lumpy universe at a very low redshift:}
\begin{eqnarray}\label{eq:averageareadistance}
\averageA{  {d}_A(z,{\n})}= \bar{d}_A(z)\left[1 +  \averageA{ \frac{\xi{\p}}{\bar{d}_{A}} } \right] = \frac{{d}^{\rm{eff}}_{H}(z) }{(1+z)} \int _{z_{\rm{cut}}}^{z}\frac{dz'}{{\sqrt {\Omega _{m}(1+z')^{3}+\Omega _{\Lambda }}}}\,,% \rightarrow \lim_{z\to0} \averageA{  {d}_A(z,{\n})} 
\end{eqnarray}
where $\Omega_{m}$ is the matter energy density parameter, $\Omega _{\Lambda } = 1 -\Omega_m$ is the energy density due to the cosmological constant and ${d}^{\rm{eff}}_{H}$ is the effective Hubble distance which captures the effect of the deformation to the observer spacetime
\begin{eqnarray}
{d}^{\rm{eff}}_{H}(z) = d_{H} \left[1 +  \averageA{ \frac{\xi{\p}}{\bar{d}_{A}} } \right]
&=& d_{H} +  \frac{5d_{H}}{ r(z) }( f(z)D(z))^2 \sum_{\ell =2}^{\ell_{\rm{max}}} (2\ell+1)
 \int \frac{d k_1}{2\pi^2} k_{1}P_{m}(k_1)
 j'_{\ell}(k_1  r(z))j''_{\ell}(k_1  r(z))\,.
% \\
 %&\stackrel{{{ z\to0}  }}{=} & d_{H} +   \frac{4}{45} d_{H}  f^2(0) \sigma^2_{R}=d_{H} \left[1+  \frac{2}{15}\averageA{\frac{ \sigma_{ij}\sigma^{ij}}{H^2_0}}\right]\,,
 \label{eq:effectiveHubblelength1}
\end{eqnarray}
Here $ d_{H}=c/H_0$ is the global Hubble distance.   {The correction given in equation \eqref{eq:averageareadistance} is due to the the parallax effect, it dominates in the very low redshift region. % This is a displacement  in the apparent positions of nearby sources as measured by an observer with non-zero peculiar velocity with respect to a local over-density.  This effect is used by Heliocentric observer to obtain the distance estimates to nearby background  stellar sources~\cite{Riess:2018byc} and even extra-galactic sources~\cite{Paine:2019vep}.  
}

 %Unlike the impact of the peculiar velocity contribution, the impact of the tidal interaction 
%Therefore, the correct model of the area distance within the $\Lambda$CDM universe is given by %In the second equality, we have taken the very low redshift limit. 

%\red{The radial velocity of a cosmologically "close" object can be approximated by

%${v_{r}=H_{0}d+v_{\rm{pec}}}$
%with contributions from both the Hubble flow and peculiar velocity terms, where $H_{0}$ is the Hubble constant and $d$ is the distance to the object.}

\section{ Cosmography in the presence of structures and consequences for Hubble rate}\label{sec:cosmography}

From figure \ref{fig:totalDA} and sub-section \ref{sec:tidaldeformation}, it is clear that inhomogeneity impacts the measurement of the area distance at about the redshift of today. In this limit, the technique of cosmography is best suited for studying the area distance~\cite{Dunsby:2015ers,Efstathiou:2021ocp}. It allows estimating the Hubble without making any assumptions about the matter content of the universe.

\subsection{Low redshift limit of the area distance and the decoupled region}

Cosmography involves expanding the area distance in Taylor series around $z =0$.  Since $d_A$ is differentiable around $z=0$, it is straight-forward to expand equation \eqref{eq:areadistance} in Taylor series around $z=0$
\begin{eqnarray}\label{eq:Taylorexp}
d^{\rm{T}}_{A}(z,{\n})  = d_{A}0) + \frac{\d d_{A}(z,{\n})}{\d z} \bigg|_{z =0} z+ \frac{1}{2}\frac{\d^2 d_{A}(z,{\n})}{\d z^2} \bigg|_{z =0} z^2 + \mathcal{O}\left(z\right) ^3\,.
\end{eqnarray}
where $d^{\rm{T}}_{A}$ denotes the area distance in the low redshift limit. 
 On the background, $d_{A}(z,{\n}) \approx \bar{d}_A(z) =  r/(1+z)$, substituting in equation \eqref{eq:Taylorexp} gives
\begin{eqnarray}
\bar{d}^{\rm{T}}_A(z) &=& \frac{z}{H_0}-\frac{1}{2}\frac{\left(3+q_0\right)}{H_0} z^2+\mathcal{O}(z)^2\,,
\label{eq:FLRWdA}
\end{eqnarray}
where  $d_{A}(0)  = 0$, $H_0$ and $q_{0} $ are the Hubble and the deceleration parameter evaluated at $z =0$ respectively.  We made use of the FLRW spacetime expression for the comoving distance to introduce $H_0$ and $q_{0} $
\begin{equation}
 \frac{\d  r}{\ dz} \bigg|_{z=0} = \frac{1}{H_0} \,, 
 \qquad \qquad \qquad
 \frac{\d^2  r}{\d z^2} \bigg|_{z =0}  =-\frac{H'(z)}{H^2}\bigg|_{z=0} =- \frac{q_{0} +1}{H_0}\,.
\end{equation}
%where we made use of equation \eqref{eq:radialxi}, its the leading order correction is given in equation \eqref{eq:redshiftperturbation}.  
Putting equation \eqref{eq:areadistance} into equation \eqref{eq:Taylorexp} we find
\begin{eqnarray}\label{eq:pertCosmography}
d^{\rm{T}}_A(z,{\n}) &=&  \frac{1}{H_0} \left[ 1 - \frac{\d \delta z}{\d z} + \left( 2+ q_0\right) \delta z \right] z
\\ \nonumber &&
+\frac{1}{2}\frac{1}{H_0}\bigg[ - \left( q_0+3\right) 
+  H_0\frac{\d^2 }{\d z^2}\left( \bar{d}_{A} \kappa\right)
- \frac{\d^2 \delta z}{\d z^2}  + 2 \left(2+q_0\right) \frac{\d \delta z}{\d z} 
+ \left( j_0-3 q_0^2 -  5q_0 -5  \right) \delta z \bigg] z^2 + \mathcal{O}(z^3)\,,
\end{eqnarray}
where we have introduced the jerk parameter  through
${H''(z)}/{H }\big|_{z=0}\equiv j_0 -q_0^2$ for completeness~\cite{Bolotin:2015dja}.  $\bar{d}_A \gamma_{ij} \gamma^{ij}$ and $\bar{d}_A\omega_{ij}\omega^{ij}$ do not contribute at this order in redshift expansion when evaluated at the redshift of today. They will obviously contribute when evaluated at different redshift position as describe in \cite{Cattoen:2007sk}.
We convert redshift derivative to radial derivative or derivative along the line of sight using Chain rule
 \begin{eqnarray}
\frac{\d \delta z}{\d z} \bigg|_{z =0} & \approx& \frac{\partial \delta z}{\partial  r} \frac{\partial  r}{\partial z} \bigg|_{z =0}  
=\frac{1}{H_0} \frac{\partial \delta z}{\partial  r}  \bigg|_{z =0} \,,\label{eq:spatialderivative1}
\\
 \frac{\d^2 \delta z}{\d z^2} \bigg|_{z =0}  &=&\bigg[\frac{\partial^2  r}{\partial z^2} \frac{\partial \delta z}{\partial  r}   +  \frac{\partial^2 \delta z}{\partial  r^2} \left(  \frac{\partial  r}{\partial z}\right)^2\bigg]_{z =0} 
 =-\frac{\left( q_{0} +1\right)}{H_0} \frac{\partial \delta z}{\partial  r} \bigg|_{z =0}    +  \frac{1}{H^2_0}\frac{\partial^2 \delta z}{\partial  r^2} \bigg|_{z =0} \ \,.
 \label{eq:spatialderivative2}
\end{eqnarray}
Putting all these together, the leading order corrections to  perturbed expression for the area distance becomes
\begin{eqnarray}\label{eq:pertburbedareadistance}
d^{\rm{Tr}}_A(z,{\n}) &\approx&    \frac{1}{H_0} \left[ 1 -\frac{1}{H} \frac{\partial \delta z}{\partial  r}  \right]_0z
+\frac{1}{2}\frac{1}{H_0}\bigg[ - \left( q_0+3\right) 
  +  \frac{1}{H} \frac{\partial^2\left( \bar{d}_{A} \kappa\right)}{\partial  r^2}
+{\left( 3q_{0} +5\right)}\frac{1}{H} \frac{\partial \delta z}{\partial  r} 
%\\ \nonumber && 
  -  \frac{1}{H^2} \frac{\partial^2 \delta z}{\partial  r^2}     \bigg]_0z^2 + \mathcal{O}(z^3)\,,\qquad
\end{eqnarray}
 In conformal Newtonian gauge, the redshift perturbation  and the lensing convergence are given by
 \begin{eqnarray}
 \delta z( r_s,{\n}) &\approx&\partial_{\|}{v\one_s}
-\frac{1}{\HH} \partial_{\|}{v\one_s}\partial^2\p{v\one_s}
-\nabla_{\bot j}\partial_{\|}{v\one_s}\int^{ r_s}_0 {( r -  r_s)}\nabla_{\bot}^i(\Phi\one + \Psi\one) \d r\,,
\label{eq:redshiftperturbation}
\\
 r \kappa( r_s,{\n})  & \approx&-\frac{1}{2}
 \nabla_{\Omega i} \int^{ r_s}_{0}\int^{ r}_{0}\nabla_{\bot }^{i}\left(\Psi\one+\Phi\one\right)\d r  \d r \,.
 \label{eq:kappa}
 \end{eqnarray}
 Details on the derivation of these expressions are given in the Appendix \ref{deviationvector}.
Since we are truncating the Taylor series expansion at second order (see equation \eqref{eq:Taylorexp}), this implies that only terms with a maximum of two LoS integrals can contribute to equation \eqref{eq:Taylorexp}
\begin{eqnarray}
\frac{\partial \delta z}{\partial  r} \bigg|_{z =0} & =&\partial_{\|}^2{v\one_o}
-\frac{1}{H_{o}} \left( \partial^2_{\|}{v\one_o}\partial^2\p{v\one_o} + \partial_{\|}{v\one_o}\partial^3\p{v\one_o}\right)\,,\label{eq:singledeltaz}
\\ 
\frac{\partial^2 \delta z}{\partial  r^2}\bigg|_{z =0}  &=& \partial_{\|}^3{v\one_o}
-\nabla_{\bot j}\partial_{\|}{v\one_o}\nabla_{\bot}^{j}(\Phi\one + \Psi\one)_{o} 
%\\ \nonumber &&
-\frac{1}{H_o} \bigg( 2\partial^2_{\|}{v\one_o}\partial^3\p{v\one_o}  
+ \partial^2_{\|}{v\one_o}\partial^3\p{v\one_o} +  \partial_{\|}{v\one_o}\partial^4\p{v\one_o}
\bigg)\,,\label{eq:doubledeltaz}
%\\ \nonumber &&
\\ 
\frac{\partial^2 \left( \bar{d}_{A} \kappa\right)}{\partial  r^2}\bigg|_{z =0}   &=&- \nabla_{\bot i}\nabla_{\bot }^{i}\left(\Psi\one+\Phi\one\right)_{o}\,.
\end{eqnarray}

It is important to understand how well the Taylor series expansion approximates the full expression for the area distance.  
On the background FLRW spacetime level, cosmography at second order in redshift is valid up to $z=0.1$. When the inhomogeneities are added,  the difference becomes
\begin{eqnarray}\label{eq:eff_dA}
\averageA{{d}^{\rm{T}}_A(z, {\n})} - \bar{d}^{\rm{T}}_{A}(z)&=& \frac{z}{H_0}\left[ \lim_{z\to 0}  \averageA{ \frac{\xi{\p}}{\bar{d}_{A}} }- \frac{1}{2}(3q_0+5) \lim_{z\to 0}  \averageA{ \frac{\xi{\p}}{\bar{d}_{A}} } z+\mathcal{O}(z)^2\right]\,.
\end{eqnarray}
Again, to arrive at this,  we assumed that the initial density field is Gaussian, which implies that the expectation values of the following terms vanish
\begin{eqnarray}
\averageA{\frac{\partial^2 \delta z}{\partial  r^2} \bigg|_{0}} = 0=\averageA{\frac{\partial^2 \left( \bar{d}_{A} \kappa\right)}{\partial  r^2}\bigg|_{z =0}  } \,.  
\end{eqnarray}
Comparing equation \eqref{eq:eff_dA} in this form to the full expression,  i.e $ \averageA{d_A(z,{\n})} - \bar{d}_A \approx \bar{d}_A \averageA{\xi_{\p}/\bar{d}_A}$ allows to focus precisely on the corrections since the convergence properties of the background is well known already.
 We show the result of computing equation \eqref{eq:eff_dA} for different smoothing scales and comparing it to the full expression (equation \eqref{eq:averageDA}) in figure \ref{fig:Error}.  
  \begin{figure}[h]
%\centering 
\includegraphics[width=80mm,height=70mm] {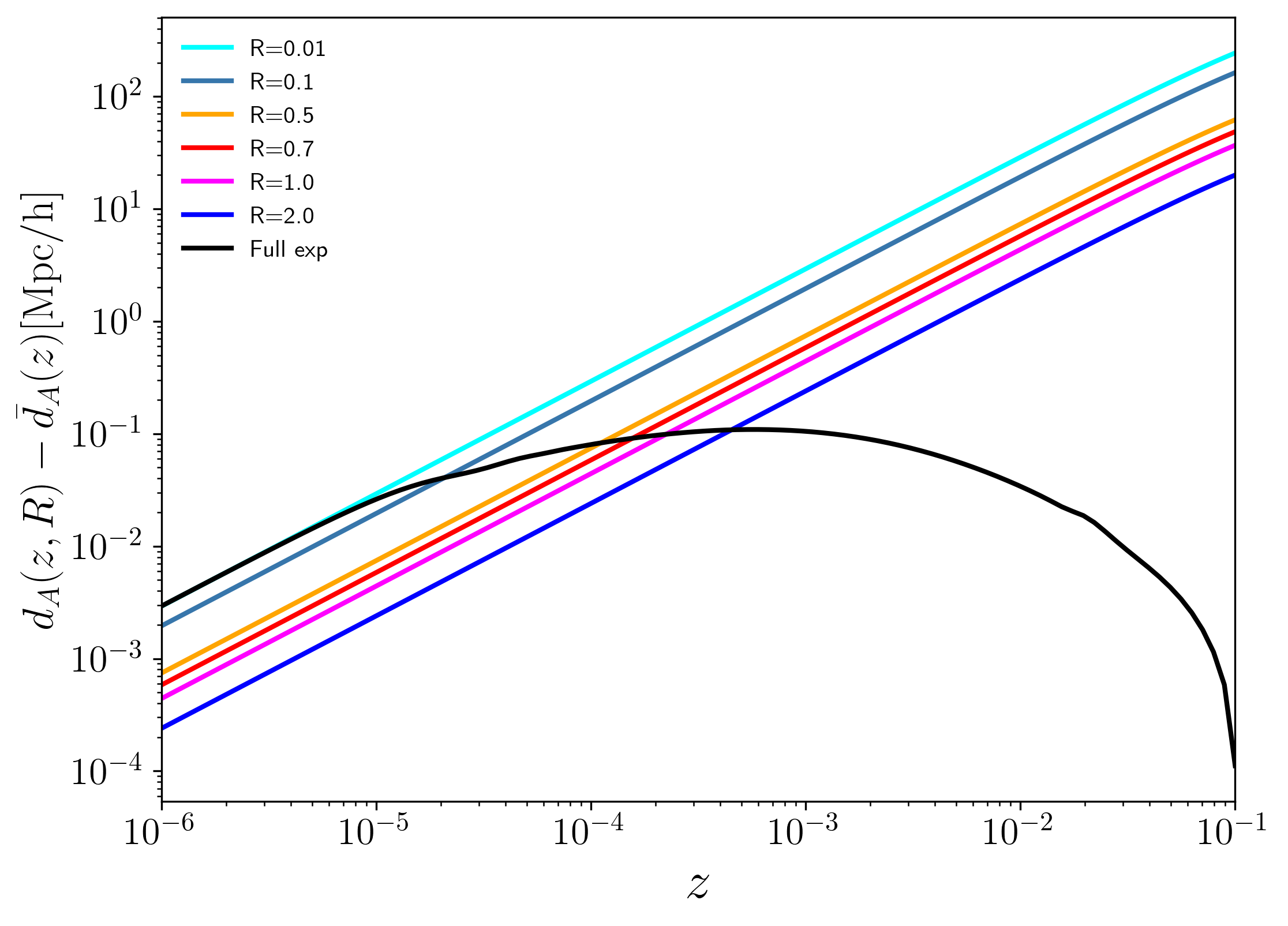}
\includegraphics[width=80mm,height=70mm] {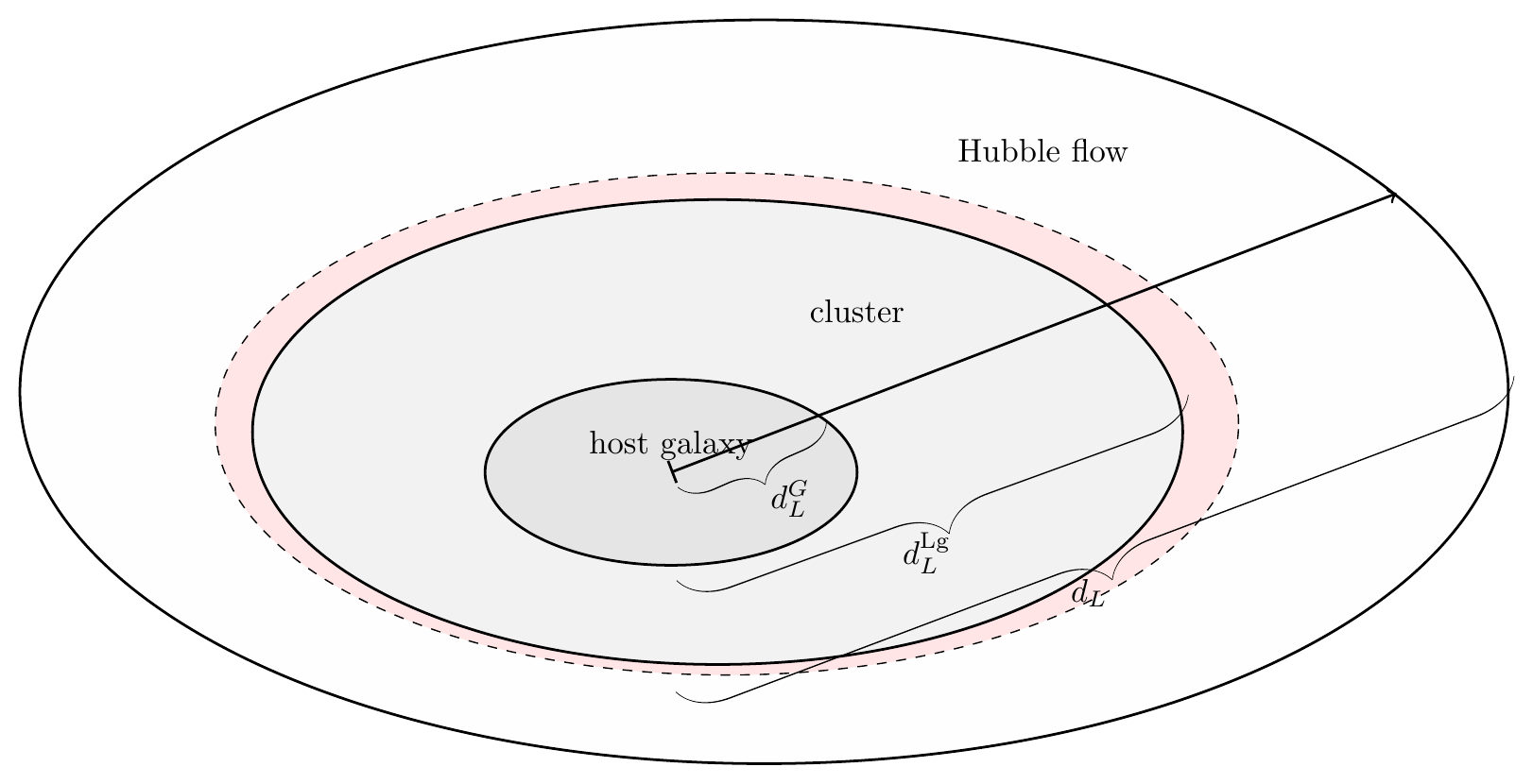}
\caption{\label{fig:Error} The left panel shows the difference between the cosmography for the area distance (equation \eqref{eq:eff_dA}) at different smoothing scales and the full expression (equation \eqref{eq:averageDA}). The unit of the smoothing scale is in Mpc/h. On the right panel, we illustrate the geometry of the problem. The observer is at the barycentre of the host galaxy for simplicity. The observer uses distance measurements to nearby structures with some of them within our local group as anchors to calibrate SNIa in the Hubble flow. The model of distance within the galaxy $d_L^{G}$ and the local group, $d_L^{\rm{Lg}}$, are not given by the background FLRW metric. We know this because in the neighbourhood of our local group, we showed that the electric part of the Weyl tensor is non-vanishing (see equation \eqref{eq:averagetidal}).
The distance within this region depends on the scalar invariant of the rate of shear deformation. The dashed circumference indicates the start of the zero-velocity surface (red shaded area).
However,  the distance  in the Hubble flow $d_L- (d_L^{\rm{G}} + d_L^{\rm{Lg}})$ can be determined using the FLRW metric. 
}
\end{figure}
%r
 It is immediately clear that we need higher-order redshift corrections to achieve convergence to a significant accuracy at about $z=0.1$. 
This an indication that the cosmography approach is less reliable at high redshift when the effects of inhomogeneities is taken in consideration.   It cannot be used for studying the expansion history of the universe. The full expression will be needed for the expansion history. However, it is valuable for constraining the Hubble rate provided the focal points in the neighbourhood of the observer are avoided, i.e set the smoothing scale at or above thee radius of the zero-velocity surface. The SNIa,  for example, only the information contained in the intercept of the Hubble diagram is needed to determine the Hubble rate and not the entire history~\cite{Riess:2021jrx}.

\subsection{Supernova peak magnitude-redshift relation}\label{sec:correctHO}

The SNIa have a consistent peak luminosity  that allows them to be used as standard candles to measure the distance to their host galaxies.  The apparent magnitude of any source, $m$,  is related to the observed flux density $F_{d_L}$  in a given spectral filter according to 
\begin{eqnarray}
m=-2.5\log _{10}\left[F_{d_{L}}\right]\,,
\end{eqnarray}
Similarly, the absolute magnitude, $M$, of the source  is defined as the apparent magnitude of the same source but measured at a  distance $D_{F}$
\begin{eqnarray}
M=-2.5\log _{10}\left[{F_{D_F}}\right]\,,
\end{eqnarray}
where $F_{D_F}$ is  called the reference flux or the zero-point of the filter~\cite{2015PASP..127..102M}. The observed flux density is related to the luminosity distance according to
\begin{eqnarray}
F_{d_L}= \left[\frac{D_{F}}{d_L}\right]^2 F_{D_{F}}\,.
\end{eqnarray}
This is the well-known  inverse square law for the source brightness. The distance modulus is defined as the difference between $m$ and $M$
\begin{eqnarray}\label{eq:distancemod}
m - M  = - 2.5 \log \left[\frac{ F_{d_L}}{F_{D_{F}}}\right] = 5 \log \left[\frac{d_{L}}{{D_F}}\right]
= 5 \log \left[\frac{d_{L}}{[\rm{pc}]}\right] - 5 \,.
\end{eqnarray}
In the last equality, we set  $D_{F} =10 \,{\rm{pc}}$ for historical reason.  In this case, it means that the absolute  magnitude  is  the apparent  magnitude  if  placed at  a distance of 10 pc. In cosmology however,  $D_{F}$ is set to  $D _{F}=1 \,{\rm{Mpc}}$  leading to 
\begin{eqnarray}
{\mu}(z,{\n})&=& m(z,{\n})-M
= 5 \log \left[\frac{d_{L}}{[\rm{Mpc}]}\right] +25 \,,% 25 + 5 \log_{10} {d}_{L}(z,{\n})\,,
\label{eq:mag-redshift-relation0}
\end{eqnarray}
where ${d}_L$ is in  the  units of ${\rm{Mpc}}$.  Again, this  implies that absolute magnitude  is the apparent magnitude of a source if it were to be measured at $1 {\rm{Mpc}}$. 
% If one requires the redshift to be an monotonically increasing function  for sources in the 
%This is very crucial in the discussion that follows. This is an important scale whose importance in cosmology is yet to be appreciated. 

The  Etherington reciprocity theorem  which holds in any metric theory of gravity in geometric optics limit allows to obtain the luminosity distance from the area distance: $d_{L}  = d_{A}(1+z)^2$:
\begin{eqnarray}\label{eq:TaylorexpdL}
d_{L}(z,{\n})  
&=&  \frac{\d d_{A}(z,{\n})}{\d z} \bigg|_{z =0} z\bigg[1+ \frac{1}{2}\left( \frac{\d^2 d_{A}(z,{\n})}{\d z^2}\left[\frac{\d d_{A}(z,{\n})}{\d z} \right]^{-1}+ 4 \right) \bigg|_{z =0}z+  \mathcal{O}\left(z\right) ^2\bigg]\,.
\label{eq:TaylorexpdLfactored}
\end{eqnarray}
We have factored out the coefficient of order one redshift expansion. Having the luminosity distance in this form is key to isolating the impact of the tidal field on the calibration of the SNIa using local distance anchors.  
Since we have already calculated the full expression for the area distance in equation \eqref{eq:pertburbedareadistance}, it is straightforward to obtain
the luminosity distance on the FLRW background
\begin{eqnarray}\label{eq:backgrounddL}
 \bar{d}_{L}(z)&=& \frac{z}{{H}_0}  +\frac{1}{2} \frac{(1-{q}_0)}{H_0} z^2 + \mathcal{O}(z^3)\,.
\end{eqnarray}
And in the presence of perturbations it is given by
\begin{eqnarray}\label{eq:inhomogenousdL}
d_{L}(z,{\n})  &=&       \frac{1}{H_0} \left[ 1 -\frac{1}{H} \frac{\partial \delta z}{\partial  r}   \right]_{0}z\bigg\{1 
+ \frac{1}{2} \bigg[(1-q_0)  +  \frac{1}{H} \frac{\partial^2\left( \bar{d}_{A} \kappa\right)}{\partial  r^2} 
  -  \frac{1}{H^2} \frac{\partial^2 \delta z}{\partial  r^2} 
  +  \frac{1}{H^2}\frac{\partial \delta z}{\partial  r} \frac{\partial^2\left( \bar{d}_{A} \kappa\right)}{\partial  r^2} 
\\ \nonumber &&
  -  \frac{1}{H^3}\frac{\partial \delta z}{\partial  r} \frac{\partial^2 \delta z}{\partial  r^2}  \bigg]_0z
  +  \mathcal{O}\left(z\right) ^2\bigg\}\,.
\end{eqnarray}
The monopole of the distance modulus(equation \eqref{eq:mag-redshift-relation0} ) with $d_L$ given by the perturbed expression for the luminosity distance (i.e equation \eqref{eq:inhomogenousdL})) is given by
 \begin{eqnarray}\label{eq:mag-redshift-relationCCHP}
\averageA{\mu}=\averageA{m}-\averageA{M}& =& 25 + 5\averageA{ \log_{10} d_{L}}\,.  %\\
\end{eqnarray}
The key job here is to evaluate the monopole of the logarithm of $d_L$, i.e $\averageA{ \log_{10} d_{L}}$. 
We start this process by takin the logarithm of $d_{L}$
\begin{eqnarray}\label{eq:log10dL}
\log_{10}\left[d_{L}(z,{\n}) \right] &=&- \log_{10}H_0+\log_{10}\left[ 1 -\frac{1}{H} \frac{\partial \delta z}{\partial  r}  \bigg|_{z =0} \right]+ \log_{10}\hat{d}_{L}(z,{\n})\,.
\end{eqnarray}
We have introduced the Hubble rate normalised luminosity distance
\begin{eqnarray}
\hat{d}_{L}(z,{\n})&=&cz\bigg\{1 + \frac{1}{2} \bigg[(1-q)  +  \frac{1}{H} \frac{\partial^2\left( \bar{d}_{A} \kappa\right)}{\partial  r^2} 
  -  \frac{1}{H^2}\frac{\partial^2 \delta z}{\partial  r^2}
  +  \frac{1}{H^2}\frac{\partial \delta z}{\partial  r} \frac{\partial^2\left( \bar{d}_{A} \kappa\right)}{\partial  r^2} 
  -  \left(\frac{1}{H}\right)^3 \frac{\partial \delta z}{\partial  r} \frac{\partial^2 \delta z}{\partial  r^2} \bigg]_{0} z
  +  \mathcal{O}\left(z\right) ^2\bigg\}\,.
\end{eqnarray}
Now, we take the  monopole of equation \eqref{eq:log10dL} 
\begin{eqnarray}\label{eq:monopoleoflog}
\averageA{\log_{10}\left[d_{L}(z,{\n}) \right] }&=&- \log_{10}H_0+\averageA{\log_{10}\left[ 1 -\frac{1}{H} \frac{\partial \delta z}{\partial  r}  \bigg|_{z =0} \right]}
+ \averageA{\log_{10}\hat{d}_{L}(z,{\n})}\,.
\end{eqnarray}
In the limit ${\partial \delta z}/(H_0{\partial  r}) \ll1$, we can evaluate the monopole of the second term by expanding in Taylor series. This is equivalent to assuming that the contribution of the anisotropic part is very small when compared to the contribution of the monopole part. After a little algebra we find
\begin{eqnarray}
\averageA{\log_{10}\left[ 1 -\frac{1}{H_0} \frac{\partial \delta z}{\partial  r}  \bigg|_{z =0} \right]}
&\approx &\log_{10} \left[1+ \frac{1}{15}  \averageA{\frac{ \sigma_{ij}\sigma^{ij}}{H^2}\bigg|_{z=0}}\right]\,.
\label{eq:monopolelog-logmonopole}
\end{eqnarray}
In the second equality, we have made use of equation \eqref{eq:averagetidal} to express the result in terms of the scalar invariant of the shear tensor.  Within the limit of our perturbation theory approximation we find that the monopole of the Hubble rate  normalised luminosity distance is given by
\begin{eqnarray}
\averageA{\log_{10}\hat{d}_{L}(z,{\n})} = \log_{10}\left[cz\bigg\{1 + \frac{1}{2}(1-q_0) z+  \mathcal{O}\left(z\right) ^2\bigg\}\right]\,.
\end{eqnarray}
In general, there could be corrections to the FLRW approximation especially at at second order in redshift. W e neglected terms such as   $\partial_{||} v\one_{s} \partial_{||} v\one_{s} $ and $\partial_{||}\Phi\one_s\partial_{||} v\one_s/\HH_s$  which contributes at second order in redshift expansion (see equation \eqref{eq:Taylorexp}). These terms will come from the second order radial perturbation and sub-dominant post-Born corrections. Some more details about  this and how it is derived may be found in \cite{Umeh:2012pn,Umeh:2014ana,Clarkson:2014pda}. We provide a general coordinate independent derivations of these equations in \cite{Umeh:2022prn}, see also~\cite{Heinesen:2020bej,Macpherson:2021gbh} for an alternative presentation.
Putting all these back into equation \eqref{eq:mag-redshift-relationCCHP} gives a smooth distance modulus
\begin{eqnarray}\label{eq:smoothmu}
\averageA{\mu}(z,H_0,q_0) &=&\averageA{m}-\averageA{M}^{R}
= 5\log_{10} \left[\frac{cz}{H_0} \bigg(1+ \frac{1}{2}(1-q_0)z + \mathcal{O}(z^2)\right] + 25\,,
\end{eqnarray}
where we have introduced a renormalised absolute magnitude  $\averageA{M}^{R}$
\begin{eqnarray}\label{eq:Mrenorm}
\averageA{M}^{R}&=& \averageA{M}+ 5\log_{10} \left[1+ \frac{1}{15}  \averageA{\frac{ \sigma_{ij}\sigma^{ij}}{H^2}\bigg|_{z=0}}\right]\,.
\end{eqnarray}
The scalar invariant of tidal deformation tensors has now been absorbed into the definition of the absolute magnitude. The right-hand side of equation \eqref{eq:smoothmu} corresponds to the background FLRW model. The effects of the inhomogeneities on the area distance at very low redshift we described in sub-section \ref{sec:allsky} or equation \eqref{eq:averageareadistance}  impacts the calibration of the absolute magnitude.  
The difference between the renormalised absolute magnitude $\averageA{M}^{R}$ and the intrinsic absolute  $\averageA{M} $ is given by
\begin{eqnarray}
\averageA{M}^{R} - \averageA{M} &=&5\log_{10} \left[1+ \frac{1}{15}  \averageA{\frac{ \sigma_{ij}\sigma^{ij}}{H^2}\bigg|_{z=0}}\right]\approx   \frac{2}{9\log10}f^2(0) \sigma^2_{R} \approx 0.12{ [\rm{mag}]}\,.
\end{eqnarray}
To evaluate this difference, we set $R = D_{F} = 1 {\rm{Mpc}}$. 
%This ensures that redshift is a monotonically increasing function in an expanding universe. 

\subsection{Calibration of the SNIa using local distance anchors}\label{sec:SHoES}

SHoES collaboration's estimates of $H_0$ relies mainly on the constraint on the  intercept of the distance modulus~\cite{Riess:2016jrr} 
\begin{eqnarray}\label{eq:intercept}
 a_b&=& -\frac{1}{5}\left( M_b+ 25 - 5 \log_{10} H_0\right)\,,
\end{eqnarray}
where $M_b$ is called standardisable absolute luminosity for the SNIa. It is calibrated using the measurement of the distance modulus to cepheids that lives in the same galaxy host with at least one SNIa
\begin{eqnarray}\label{eq:standardizable}
M_b = m_{b,\rm{SNIa}} -\mu_{0,\rm{ceph}}\,, %= M -\mu_{0,\rm{ceph}}\,,
\end{eqnarray}
where $ m_{b,\rm{SNIa}}$ is the apparent magnitude of a nearby Supernova with cepheids within the same host as the cepheids and $\mu_{0,\rm{ceph}}$ is an independent distance modulus to the cepheids. The SHoES collaboration uses parallax methods to estimate the distance to the Milky Way Cepheids~\cite{Riess:2018byc}. From this distance, the absolute magnitude of cepheids is determined.
The collaboration has since made use of different geometrical distance estimates such as the distance to the NGC 4258 obtained by modelling of the water masers in the nucleus of the galaxy orbit about its supermassive black hole~\cite{Reid:2019tiq} and the distance to the Large Magellanic Cloud obtained from the dynamics of the detached eclipsing binary systems~\cite{2019Natur.567..200P} to calibrate $M_b$.

The Hubble rate is obtained from equation \eqref{eq:intercept}
\begin{eqnarray}\label{eq:FLRW_intercept}
\log_{10} H_0= \frac{ M_b+25  + 5a_b}{5} \,,%=  \frac{ M- \mu_{0,\rm{ceph}}+25  + 5a_b}{5 }\,,
\end{eqnarray}
where $a_b$ is found  from fitting the intercept of the distance modulus of the SNIa which constrains $M_b + 5a_b$ %and from equation \eqref{eq:lum-red-relation}, we find
\begin{eqnarray}\label{eq:ab}
a_b &=& \log_{10}\bigg\{ cz \left[ 1 + \frac{1}{2}(1-q_0) z\right]\bigg\}- 0.2 {M_b}\approx\log_{10}  cz  - 0.2 {M_b}\,.
\end{eqnarray}
%obtain the constraint on the  intercept from the distance modulus
Starting  from equation \eqref{eq:smoothmu} for  a smoothed inhomogeneous model for the luminosity distance we find that the monopole of the intercept is given by
 \begin{eqnarray}\label{eq:average_intercept2}
\averageA{a_b}
% &=& -\frac{1}{5}\left[ \averageA{M}+ 5 \log_{10} \left[1+ \frac{1}{15}  \averageA{\frac{ \sigma_{ij}\sigma^{ij}}{H_0^2}}\right]+ 25 - 5\log_{10} H_0\right]\,, \\
 &=& -\frac{1}{5}\left[\averageA{M}^{R}+ 25 - 5\log_{10} H_0\right]\,,
\end{eqnarray}
where  $\averageA{M}^{R}$ is given in equation \eqref{eq:Mrenorm}. The Hubble rate  and the intercept are the same as in equation \eqref{eq:FLRW_intercept} and \eqref{eq:ab} respectively with $M_b $ replaced by $\averageA{M}^{R}$.

%\subsection{The Carnegie-Chicago Hubble Program}
The Carnegie-Chicago Hubble Program, CCHP, uses TRGB to calibrate local SNIa samples. 
\begin{eqnarray}\label{eq:absMag}
M_{i} = m_{i} - \mu_{0 i}^{\rm{TRGB}}\,,
\end{eqnarray}
where $m_{i}$ is the apparent magnitude of the peak of the SNIa light curve from a given subsample that contains at least one TRBG, $\mu_0^{\rm{TRGB}}$ is the true calibrator distance modulus. Once the SNIa absolute magnitude is calibrated using a subsample of the very nearby supernova according to equation \eqref{eq:absMag}, then magnitude-redshift relation based on the background FLRW  spacetime is used for the rest of the samples ($0.03\le z\le 0 .4$) to perform a likelihood inference for  $H_0$ using equation \eqref{eq:smoothmu}. CCHP sets  $q_{0} = -0.53$ to determine the Hubble rate to be $H_0 = 69.6\pm1.9\,\, {\rm{km/sec/Mpc}}\,$.

%We will now show how the effect of the tidal deformation of the local spacetime is captured by $ \mu_{0 i}^{\rm{TRGB}}$ in equation \eqref{eq:absMag}.  

%%%%%%%%%%%%%%%%%%%%%%%%%%%%%%%%%%%%%%%%%%%%%%%%%%%%%%%%%%%

\subsection{The  Supernova absolute magnitude tension}\label{sec:SN1Amagtension}

The inverse distance ladder technique could be used to estimate the absolute magnitude of SNIa samples. It assumes the FLRW model for distances and uses the sound horizon scale at the surface of last scattering, $r_\star$ as an anchor. The CMB constraint on $r_\star$ is used to calibrate the SNIA samples~\cite{Camarena:2019rmj,Camarena:2021jlr}. 
For the Pantheon SNIa peak magnitudes with Planck constraint on $r_\star$, the absolute magnitude is found to be
$
M^{\rm{P18}} = - 19.387\pm 0.021\,\, {\rm{mag}} 
$~\cite{Camarena:2019rmj,Efstathiou:2021ocp}.
Using the parallax measurements of Milky Way Cepheids provided by the SHoES collaboration, \cite{Efstathiou:2021ocp} estimates the absolute magnitude of the Pantheon SNIa peak magnitudes and found
$
M^{\rm{E21}} = -19.214\pm 0.037\,\, {\rm{mag}}.
$
The difference between $M^{\rm{P18}} $ and $M^{\rm{E21}} $ gives the so-called supernova absolute magnitude tension 
\begin{eqnarray}\label{Mr-Mp}
M^{\rm{E21}} - M^{\rm{P18}}  = 0.173 \pm 0.04 \,\,  {\rm{mag}}\,.
\end{eqnarray}
\begin{figure}[h]
%\centering 
\includegraphics[width=90mm,height=65mm] {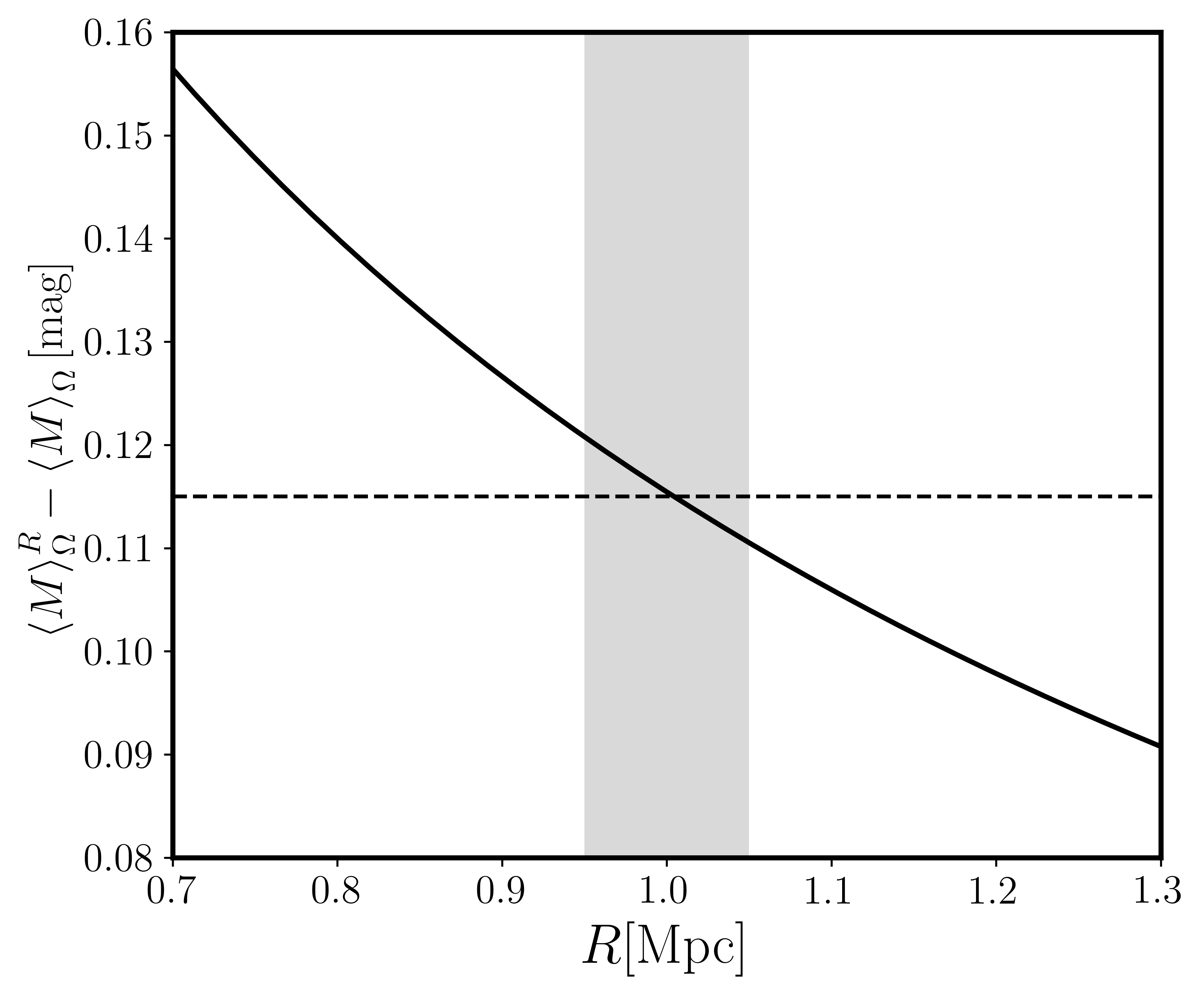}
\caption{\label{fig:Absolutemag}  
The difference between $\averageA{M}^{R} $  the intrinsic absolute magnitude as function of the smoothing scale.  The vertical grey area is  about 5\%  constraint on the radius of the surface  of the  zero-velocity that separates our local group from the Hubble flow \cite{1999A&ARv...9..273V,Li:2007eg,Karachentsev:2008st}.
 }
\end{figure}
Given that the inverse distance ladder technique assumes a model which does not include the tidal term we describe here, we associate the absolute magnitude it determines with $\averageA{M}$. The local cosmic distance ladder technique does not assume any spacetime model.  It essentially ensures the consistency in the use of the distance modulus formula ($\mu = m - M$),  hence, we associate the absolute magnitude that arises from this process to $\averageA{M}^R$. 
Therefore, for the Pantheon SNIa sample, we have 
\begin{eqnarray}\label{eq:Mdiff}
M^{\rm{E21}} - M^{\rm{P18}} =\averageA{M}^{R} - \averageA{M} &=&  \frac{1}{3\log10} \frac{ \sigma_{ij}\sigma^{ij}}{{H}^2}\bigg|_{z=0}  =   \frac{2}{9\log10}f^2(0) \sigma^2_{R}\,.
\end{eqnarray}
 We show in figure \ref{fig:Absolutemag} the difference between the absolute magnitudes calculated using equation \eqref{eq:Mdiff}. With the smoothing scale of $R=D_{F}=1.0 {\rm{Mpc}}$, we find about $0.12 [{\rm{mag}} ]$ difference between the  inverse distance ladder  predication and the local measurements.

 \subsection{Standard ruler:   Baryon acoustic oscillation}\label{sec:BAO}

 Baryon acoustic oscillation (BAO) is sensitive to the Hubble rate and the area distance through the parallel and orthogonal distortions in the separation between galaxies $\alpha_{\p}$ and $\alpha_{\bot}$ respectively. $\alpha_{\p}$ and $\alpha_{\bot}$ are known as the Alcock-Paczynski parameters~\cite{Alcock:1979mp}.  The connection between Alcock-Paczynski parameters and the Hubble rate/area distance depends on a model.  Assuming exact cosmological principle, a set of possible models reduces to the FLRW models
 \begin{eqnarray}\label{eq:BAOparams}
\bar{\alpha}_{\p} = \frac{{H}^{\rm{fid}}}{{H}} \,, \qquad\qquad \bar{\alpha}_{\bot }  = \frac{\bar{d}_{A}}{{d}^{\rm{fid}}_{A}}\,.
\end{eqnarray}
 where ${H}^{\rm{fid}}$ and ${d}^{\rm{fid}}_{A}$ are the fiducial Hubble rate and area distance respectively. They are used to estimate the separation between galaxies before
 ${H}$ and $ \bar{d}_{A}$ are adjusted to obtain the best-fit to the observed data.  The monopole  of the two-point correlation function constrains $\alpha = \alpha_{\bot}^{{2}/{3}}\alpha{\p}^{{1}/{3}}=D_{V}(z) = \left[  z D_{M}^{2}(z) /H(z)\right]^{1/3}$~\cite{Anderson:2013oza,Beutler:2014yhv}, where $D_{M}$ is the comoving distance.  The quadrupole moment is most sensitive  to the  Alcock-Paczynski ratio $\epsilon = \alpha_{\bot}/\alpha_{\p}$~\cite{Nadathur:2019mct}. 
It is possible to generalise equation \eqref{eq:BAOparams} beyond the exact cosmological principle limit to allow for stochastic inhomogeneities.  In this case, the radial  and orthogonal distance can be written as: $r{\p} = \bar{r}{\p} + \delta r  = \bar{r}{\p} + \delta z/H$ and ${r}^A_{\bot} = \bar{r}^A_{\bot} + \delta r^A_{\bot}$, then the modified Alcock-Paczynski parameters become
\begin{eqnarray}
\alpha_{ \p} &=& \frac{\partial r_{\p}}{\partial r^{\rm{fid}}_{\p }} =  \frac{\partial r_{\p}}{\partial z}\frac{\partial z}{\partial r^{\rm{fid}}_{\p}} \approx
 \frac{H^{\rm{fid}}}{H}\left[1+\frac{1}{H}\frac{\partial \delta z}{\partial r} + \mathcal{O}(\delta z)\right]\,, \label{eq:alphaparallel}
\\
\alpha_{\bot}&=& \frac{ \partial r_{\bot}^A}{\partial \theta} \frac{\partial \theta}{\partial^B r^{\rm{fid}}_{\bot }}  \approx \frac{d_{A}}{d^{\rm{fid}}_{A}} \,.
\label{eq:alphabot}
\end{eqnarray}
We can take the direction average of equations \eqref{eq:alphaparallel} and \eqref{eq:alphabot} and set the result equal to equation \eqref{eq:BAOparams}   
\begin{eqnarray}
\averageA{\alpha_{ \p} } = \bar{\alpha}_{\p} \qquad  {\rm{and}} \qquad \averageA{\alpha_{\bot}} = \bar{\alpha}_{\bot }\,.
\end{eqnarray}
Keeping the fiducial cosmology fixed in both equations  \eqref{eq:alphaparallel} and \eqref{eq:alphabot}  and equation \eqref{eq:BAOparams}  and evaluating the all sky average, we find the effective Hubble rate 
\begin{eqnarray}
{H^{\rm{eff}}}&=&  H\left[1-\frac{2}{15}\averageA{\frac{ \sigma_{ij}\sigma^{ij}}{H^2}}\right]\,.
%\qquad {\rm{and}} \qquad
%  {d}^{\rm{eff}}_A=\bar{d}_A(z)\left[1 +  \averageA{ \frac{\xi{\p}}{\bar{d}_{A}} } \right]\,.
  \label{eq:Hubblevsalphap}
 \end{eqnarray}
 and the  area distance is given in equation \eqref{eq:averageareadistance}.
 By setting  the all sky average of equations  \eqref{eq:alphaparallel} and \eqref{eq:alphabot}  to equation \eqref{eq:BAOparams}, we are asking what will be the inferred Hubble rate and the area distance to a source at a fixed redshift if the inhomogeneous universe  described by the metric given in equation \eqref{eq:metric} is interpreted using the background FLRW background spacetime.  
     \begin{figure}[ht]
%\centering 
\includegraphics[width=100mm,height=75mm] {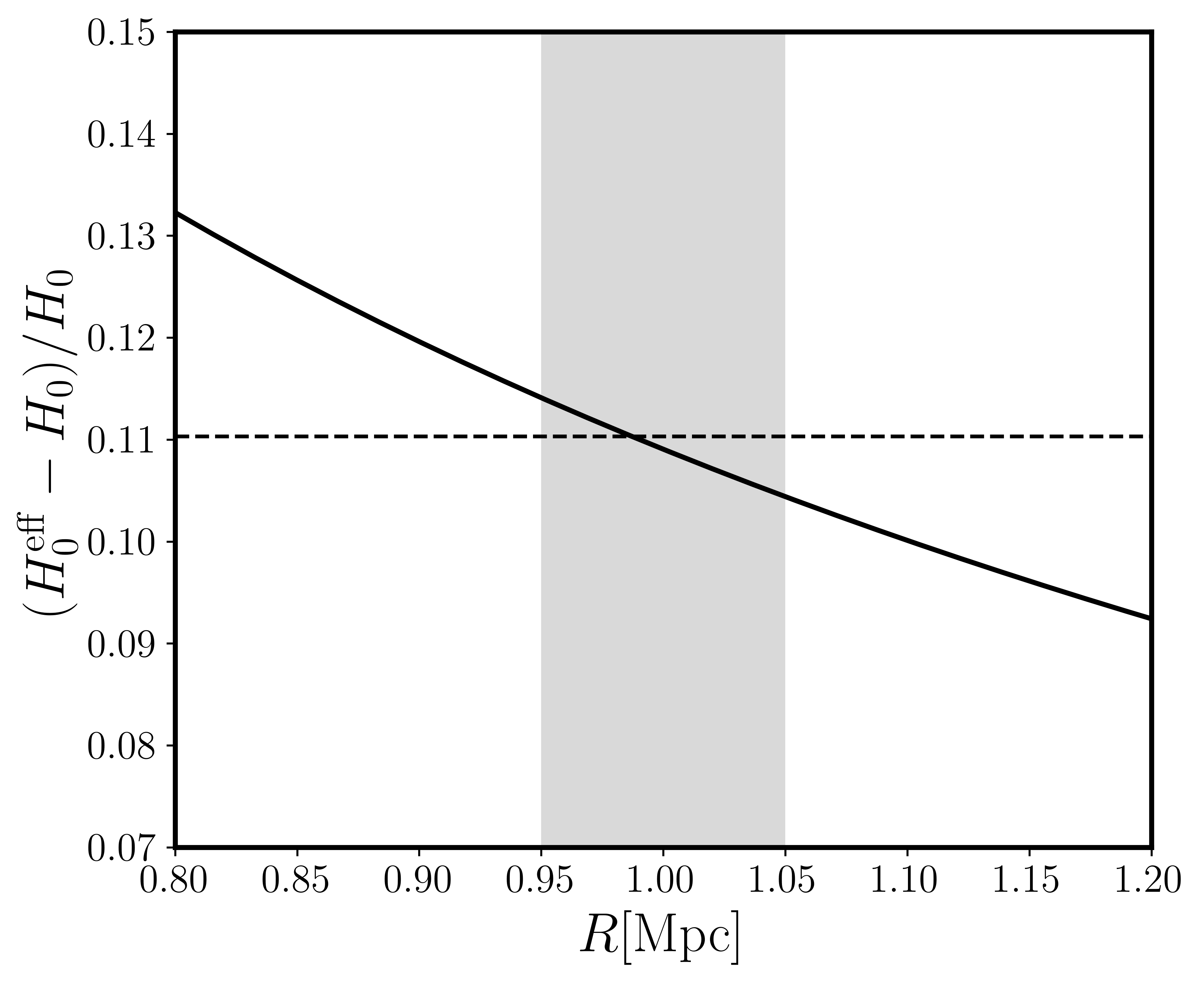}
\caption{\label{fig:Handq}   We show the dependence of the fractional difference between the effective  and global Hubble rates.
Again the vertical grey area is a constraint on the radius of the  zero-velocity surface that separates our local group from the Hubble flow \cite{1999A&ARv...9..273V,Li:2007eg,Karachentsev:2008st}.  The vertical axis is an absolute value of the difference. 
}
\end{figure}
The fractional change in the area distance has already been discussed in sub-section \ref{sec:allsky} and the key results shown in figure \ref{fig:totalDA}, hence we focus on the fractional difference between the effective and the global Hubble rate:
 \begin{equation}\label{eq:Hubblediff4}
\frac{{H^{\rm{eff}}_0} - {H_0}}{ {H_0} } = -\frac{4}{45}f^2(0) \sigma^2_{R}\,.
\end{equation}
We plot equation \eqref{eq:Hubblediff4} in figure \ref{fig:Handq} as a function of the smoothing scale. Note that $H_0$ is the Hubble rate measured by the local distance ladder technique.
We find about $(8-12)\%$ change to the Hubble rate when the tidal field is smoothed at the zero-velocity surface of our local group. 

%\cite{Dainotti:2022bzg,Dainotti:2021pqg,}

\section{Conclusion}\label{sec:conc}

We showed how to derive a very concise expression for the area distance in a lumpy universe.  
%This new approach makes it easier to analytically isolate and study the impact of the local inhomogeneities on distance measurements. 
This approach makes it easier and straightforward  to analytically isolate and  estimate the impact of the general relativistic corrections to the multipole moments of the area distance when compared to the perturbation of the full null focusing equation~\cite{Umeh:2012pn,Umeh:2014ana}
To obtain this simplified view-point, we made use of the general coordinate transformation to derive a relationship between the source position and the image position. The relative difference between these positions can also be expressed in terms of the deviation vector when the image position is identified as the position of the central ray.
We showed how to perform an irreducible decomposition of the Jacobian of the transformation with respect to the line of sight direction of the observer. {This decomposition scheme follows the ``Cosmic Ruler" decomposition approach introduced in \cite{Schmidt:2012ne}}. The orthogonal component of the deviation vector leads to the traditional Weak gravitational lensing parameters while the contribution of radial component in the Jacobian projected unto the screen space leads to the {radial lensing correction/effect} that we studied further. On the surface of constant redshift, the radial component is sourced by the perturbation in the observed source redshift. {On the constant redshift surface, the radial lensing effects is known as the Doppler lensing effects and we showed that the leading order part is the parallax effect.  That is the displacement in the background source position  due to the relative velocity of the observer. }
We showed how to evaluate the impact of the {parallax effect } on the monopole of the area distance and the distance modulus.  

{In order to calculate correctly the impact of the parallax effect on the monopole of the area distance, we needed a model of the heliocentric peculiar velocity. The heliocentric peculiar velocity cannot be calculated within our cosmological perturbation theory scheme~\cite{Macaulay:2011av}. Therefore, to maintain consistency, we positioned the observer at the barycenter of our local group, this allows us to use the techniques developed in the study of the CMB temperature anisotropies to perform a full sky spherical harmonic decomposition of the correction to the area distance}.  We discussed how to perform this calculation in greater detail in Appendix \ref{sec:harmonicsdecomp}.  {The benefit of this approach is that the contribution of the heliocentric peculiar velocity can easily be included by boosting to the heliocentric frame when the correct heliocentric peculiar velocity is determined.  The impact of the heliocentric peculiar velocity on the luminosity distance was studied in \cite{Kaiser:2014jca}.}

{Similarly, when calculating the area distance using the general coordinate transformation approach or using the focusing equation (Sachs equation or the geodesic deviation equation of the Jacobi equation) we are in principle assuming that the observer and the source are connected by a one-parameter family geodesics.  This does not hold in  our universe  because there exists a conjugate point at the boundary of our local group~\cite{Nicolaescubook,Dalal:2020mjw}.  The  focusing equation or the coordinate system induced by a one-parameter family of geodesics breaks down at a conjugate point~\cite{Witten:2019qhl}. Therefore to go beyond the conjugate point, another set of coordinate systems needs to the introduced. 
In order to avoid these complications, we analytically continued the monopole of the area distance as a function of the cosmological redshift beyond the conjugate point and then use a top-hat window function to smooth over divergent modes at  the scale $R$. We showed that the smoothing scale $R$ is given by the radius of the zero-velocity surface. What this means is that we can trust the perturbation theory prediction up the scale $R$. 
Using the constraint on the cosmological parameters given by the Planck's experiment~\cite{Aghanim:2018eyx}, we showed that the contribution of the inhomogeneities to the background area distance(or the the luminosity distance) at low redshift ($z_{\rm{cut}} \le z \le 0.001$)  could range from (4-12)\%. 
Here the $z_{\rm{cut}}$ is determined by the radius of the zero-velocity surface(see equation \eqref{eq:horizon} for details). } %The redshift fails to the monotonic below $z_{\rm{cut}}$.}
%This is because with the smoothing scale at the  % converging towards the centre of a local over-density. 

{In addition,} we studied the consequences of the modification to the monopole of the area distance in the low redshift limit on the cosmic distance ladder or {the first rung of the distance ladder}. We showed that the distance to nearby sources are impacted by the tidal deformation. The distance to the anchors  used for the calibration of the absolute magnitude of the SNIa  are located in the region where the impact of the tidal deformation cannot be neglected.  Also,  since distance is a two-point function, even sources far away in the Hubble flow are affected {via the calibration of the absolute magnitude}. We showed how the impact of the tidal deformation could bias the calibration of the SNIa absolute magnitude and we argue that it provides a natural explanation to the SNIa absolute magnitude tension. That is it explains the difference between the SNIa absolute magnitude determined using local distance anchors and the absolute magnitude obtained by inverse distance ladder with the proper length of the sound horizon at the last scattering surface as an anchor.
Using the Alcock-Paczynski parameters as observables we showed that the inferred Hubble rate from the BAO measurement under-estimates the Hubble rate of an expanding universe by about (8-12)\%. This is exactly the amount required to resolve the Hubble tension~\cite{DiValentino:2021izs}.  

{Finally, we neglected the impact of the heliocentric peculiar velocity on the luminosity distance or the apparent magnitude. It is important to note that including this coordinate dependent correction will not impact our result on the supernova absolute magnitude tension or the Hubble tension.  We expect that  the heliocentric peculiar velocity will impact the position of the intercept of the Hubble diagram. We shall study this in detail elsewhere. }

\section*{Acknowledgement}
I benefited immensely from discussions with Chris Clarkson, Robert Crittenden, Emir Gumrukcuoglu, Pierre Fleury, Asta Heinesen,  Mathew Hull, Antony Lewis, Kazuya Koyama and  Daniela Saadeh. The works by G.F.R. Ellis on this topic have been very crucial. I am supported by the 
UK Science \& Technology Facilities Council (STFC) Consolidated Grants Grant ST/S000550/1 and the South African Square Kilometre Array Project.
The perturbation theory computations in this paper  were done with the help of tensor algebra software xPand \cite{Pitrou:2013hga} which is based on xPert~\cite{Brizuela:2008ra}.
%I made use of COLOSSUS  developed by Benedikt Diemer for computations involving the dark matter halo profiles~\cite{Diemer:2017bwl}. 

%and xLightcone {\tt{https://github.com/Obinna/xLightCone}}.

%\newpage

\section*{Data Access Statement}

{No new data were generated or analysed in support of this research}
 
 \appendix
\section{ Physical Position from perturbed background}\label{deviationvector}

We derive the full expression for the source position by solving the photon geodesic equation
 \begin{eqnarray}\label{eq:geodesiceqn}
\hat{k}^b\hat{\nabla}_b\hat{ k}^a=0\,.
\end{eqnarray}
 Following Sachs and Wolfe \cite{Sachs:1967er}, the conformal transformation  $\hat{g}_{ab}\rightarrow g_{ab}$  maps the null geodesic equation associated with the physical metric $\hat g_{ab}$ (perturbed FLRW metric) to a null geodesic equation derived from  a perturbed Minkowski metric $ g_{ab}$. Both metrics are related according to $g_{ab}=a^{-2}\hat g_{ab}$. The contravariant photon 4-vector transforms as $\hat k^b=a^{-2}k^b$ while the covariant photon 4-vector transforms as $\hat k_a=k_a$.  The affine parameters  transform as $\d \hat \lambda \rightarrow \d \lambda = a^{-2} \d \hat{\lambda}$. A given time-like  4-velocity, $\hat{u}^a$ on the physical spacetime, it transforms as $\hat u^a=a^{-1}u^a\Leftrightarrow \hat{u}_a=au_a$.  The photon energy transforms as $\hat E= -\hat u_b\hat k^b = -a^{-1} \,u_b k^b=a^{-1} E$. 
 
 This implies that we can compute the photon energy on a perturbed Minkowski space-time and simply multiply the conformal factor to obtain the corresponding expression in the physical spacetime. Similarly, we can compute the source position using the perturbations of the Minkowski metric, then multiply by the scale factor to obtain the equivalent in an expanding spacetime. The solution to the geodesic equation on a perturbed Minkowski metric is given by \cite{Umeh:2012pn}
  \begin{eqnarray}
  \delta\one k^0&=&-2(\Phi\one_s-\Phi\one_o) +2\int_{\lambda_o}^{\lambda_s}{\Phi\one}'\d\lambda\,,\\
\delta\one k\p &=&  -2 \int_{\lambda_o}^{\lambda_s}{\Phi\one}'\d\lambda\,,\\
\label{eq:solngeodesicradial}
\delta\one k_{\bot }^i &=& - 2\int_{\lambda_o}^{\lambda_s} \nabla^i_{\bot }{ \Phi\one}\d\lambda\,.
  \end{eqnarray}
At second order we considered only the dominant terms. 
\begin{eqnarray}\label{eq:transverse1}
 \delta\two k^0&\approx&-2\Phi\two_s +\int_{\lambda_o}^{\lambda_s}\left({\Phi\two}'+{\Psi\two}'\right)\d\lambda+8\int
\int\frac{(\lambda_o-\tilde{\lambda})}{(\lambda_o-\lambda)}\nabla_{\bot}^i{\Phi\one}'(\tilde{\lambda})\int\nabla_{\bot i}{\Phi\one}(\tilde{\lambda})\,,\label{eq:secondk0}
\\
\delta\two k_{\p}&\approx&- \int_{\lambda_o}^{\lambda_s}\left( {\Phi\two}' +{\Psi\two}'\right)\d\lambda+8
\int \nabla_{\bot }^i{\Phi\one}\int\nabla_{\bot i}{\Phi\one}(\tilde{\lambda})
% \nonumber \\ &&
-\int\int\frac{(\lambda_o-\tilde{\lambda})}{(\lambda_o-\lambda)}\nabla_{\bot }^i{\Phi\one}'(\tilde{\lambda})\int\nabla_{\bot i}{\Phi\one}(\tilde{\lambda})
\\ 
\delta\two k_{\bot}^i &\simeq& - \int_{\lambda_o}^{\lambda_s} \nabla^i_{\bot } \left(\Phi\two+ \Psi\two\right)\d\lambda
\\  &&
+
2\int_{\lambda_o}^{\lambda_s}\left[\int_{\lambda_o}^{\lambda}\nabla_{\bot}^j
 \left(\Phi\one(\tilde{\lambda})+ \Psi\one(\tilde{\lambda})\right)\d\tilde{\lambda}\int_{\lambda_o}^{\lambda}\frac{\tilde{ r}}{ r}\nabla_{\bot j}
 \nabla_{\bot}^i \left(\Phi\one(\tilde{\lambda})+ \Psi\one(\tilde{\lambda})\right)\d\tilde{\lambda}\right]\d\lambda\,.
  \label{eq:transverse2}
\end{eqnarray}
 The full expression is given in \cite{Umeh:2012pn}.
The perturbed photon trajectory is obtain by integrating equations \eqref{eq:transverse1} and \eqref{eq:transverse2}
\begin{eqnarray}
\delta\one {\eta}(\bar{\eta}_s,{\n}) &=& 2\int_{0 }^{r_s}(r_s-r) {\Phi\one}'\d r
\label{eq:perturbedpostion1}
\\
\delta \one x\p(\bar{\eta}_s,{\n})
&\approx&  - 2\int^{r_s}_{0}\Phi\one\d r \,,
\\
\delta\one x^i_{\bot}(\bar{\eta}_s,{\n}) &
=&\int^{ r_s}_0 {( r -  r_s)}\nabla_{\bot}^i(\Phi\one + \Psi\one) \d r\,, 
\end{eqnarray}
where we have defined $r = \eta_o - \bar{\eta}$.
At second order, we need only the radial component
\begin{eqnarray}
\delta\two x_{\p}(\lambda_s,{\n}) &=& - \int^{r_s}_{0}\left(\Phi\two + \Psi\two\right)\d r+8 \int_{0}^{r_s} \d r' \int_{0}^{r'} \d r'' (r_s -r') 
 \nabla_{\bot i} \Phi\one(r'{\n}) \nabla_{\bot} ^i \Phi\one(r''{\n})
 \\ \nonumber &&
 +8\int_{0}^{r_s} \d r\int_{0}^{r} \d r' \int_{0}^{r} \d r'' \frac{r''}{r} 
 \nabla_{\bot i} \Phi\one(r'{\n}) \nabla_{\bot} ^i \Phi\one(r''{\n}) \,.
\end{eqnarray}
Together, we find that the source position as function of the Minkowski background spacetime is given by
\begin{eqnarray}\label{eq:pertposition}
%\lambda({\eta},\bar{\x})  &=& {\eta} + \delta \lambda({\eta},\bar{\x}) + \delta^2 \lambda({\eta},\bar{\x})+  \mathcal{O}(\epsilon^3)\,,
%\label{eq:pertconformaltime}\\
x^i(\bar{\eta},\bar{\x}) &=& \bar{x}^i(\bar{\eta}) +\delta\one x^i(\bar{\eta},\bar{\x}) +\frac{1}{2}\delta\two x^i(\bar{\eta},\bar{\x})+ \mathcal{O}(\epsilon^3)\,,
\end{eqnarray}
where $\bar{\eta}$ and $ \bar{x}^i$ are the background conformal time and spatial coordinate position respectively. Up to this point, the perturbations live on the background spacetime, but, light rays travel on physical spacetime, not on the background spacetime. Firstly, we need to express equation \eqref{eq:pertposition} in terms of the affine parameter of the full spacetime. On the background spacetime, we have 
\begin{equation}
\frac{\d\bar{\eta}}{\d\bar{\lambda}} =1\,,\qquad\frac{\d \bar{x}^i}{\d\bar{\lambda}} = -n^i\,.
\end{equation}
Again $n^i$ is the LoS direction vector. 
Integrating these two equations from the observer to the source give: $\bar{x}^i = -n^i(\bar{\lambda}_s-\lambda_o) = n^i r$ with $ r = (\lambda_o - \bar{\lambda}_s)=  (\eta_o - \bar{\eta}_s)$.
Given that $\lambda = \bar{\lambda} + \delta\one \lambda + \delta\two \lambda/2$ we re-map  $\bar{x}^i$  to the affine parameter associated with the perturbed Minkowski spacetime 
\begin{eqnarray}\label{eq:timemap}
\bar{x}^i(\bar{\eta}) \rightarrow \bar{x}^i (\lambda)-\delta\one \lambda \frac{\d x^i}{\d\lambda}\bigg|_s-\frac{1}{2}\left(\left(\delta\one \lambda\right)^2\frac{\d^2 x^i}{\d\lambda^2}\bigg|_s+\delta\two\lambda \frac{\d x^i}{\d\lambda}\bigg|_s\right)+ \mathcal{O}(\epsilon^3)\,.
\end{eqnarray}
Since $n^i$ is constant on the background, its acceleration vanishes
${\d^2 x^i}/{\d\lambda^2}|_s  = 0$, therefore
 \begin{eqnarray}
\bar{x}^i(\bar{\eta}) \rightarrow  r n^i+\delta\one \lambda n^i+\frac{1}{2}\delta\two \lambda n^i+ \mathcal{O}(\epsilon^3)\,.
 \end{eqnarray} 
This implies that the perturbed position given in equation (\ref{eq:pertposition}) may be written in terms of $\lambda_s$ by replacing  $\bar{x}^i({\eta})$ 
 \begin{eqnarray}\label{eq:pertpositionsub}
x^i (\lambda,{\n})&=& \bar{x}^i(\lambda) + \left[\delta\one x^i + {n^i_s}\delta\one \lambda\right]
+\frac{1}{2}\left[\delta\two x^i +n^i_s\delta\two \lambda\right]\,.
%\\
% &=&\bar{x}^i(\lambda) + \left[\delta\one x^i_{\bot} + {n^i_s} \left( \delta\one x_{\p} - \delta\one \lambda\right)\right]+\frac{1}{2}\left[\delta\two x^i_{\bot} + n^i_s\left( \delta\two x_{\p} - \delta\two \lambda\right)\right]
\end{eqnarray}
Similarly, at first order, we re-map the conformal time and position
\begin{eqnarray}
 \delta\one \lambda(\bar{\eta},\bar{\x}) &\approx&\delta\one \lambda({\lambda},{\n})-\Delta\one x_{\p}(\bar{\eta},\bar{\x}) \partial_{\p} \delta\one \lambda(\bar{\eta},\bar{\x}) -\delta\one x_{\bot}^i(\bar{\eta},\bar{\x})\nabla_{\bot i} \delta\one \lambda(\bar{\eta},{\n})\,,\label{eq:pertlambda}
 \\
  \delta\one  x_{\p}(\bar{\eta},\bar{\x}) &\approx&\delta\one x_{\p}({\lambda},{\n})- \Delta\one x_{\p}(\bar{\eta},\bar{\x}) \partial_{\p} \delta\one x_{\p}(\bar{\eta},\bar{\x}) -\delta\one x_{\bot}^i(\bar{\eta},\bar{\x})\nabla_{\bot i} \delta\one x_{\p}(\bar{\eta},{\n})\,,\label{eq:peralongpos}
  \\
  \delta\one  x^i_{\bot}(\bar{\eta},\bar{\x})  &\approx& \delta\one x^i_{\bot}(\lambda,{\n})- \delta\one x_{\bot}^i(\bar{\eta},{\n})\nabla_{\bot i} \delta\one  x^i_{\bot}(\bar{\eta},{\n})\,.
\end{eqnarray}
Again, we focused on the dominant terms only. Substituting these terms in equation \eqref{eq:pertpositionsub} gives
 \begin{eqnarray}\label{eq:trajectory3} 
  x^i (\lambda,{\n})&=&\bar{x}^i(\lambda) + \left[\delta\one x^i_{\bot} + {n^i_s} \left( \delta\one x_{\p} + \delta\one \lambda\right)\right]
+\frac{1}{2}\bigg[ \delta\two x^i_{\bot} 
 -2 \delta\one x_{\bot}^i\nabla_{\bot i} \delta\one  x^i_{\bot}
   \\ \nonumber &&
 + n^i_s\left( \delta\two x_{\p} + \delta\two \lambda -2\Delta\one x_{\p}\left( \partial_{\p} \delta\one \lambda +  \partial_{\p} \delta\one x_{\p}\right)-2\delta\one x_{\bot}^i\left(\nabla_{\bot i} \delta\one \lambda+\nabla_{\bot i} \delta\one \lambda\right)\right)\bigg]\,.
 \end{eqnarray}
 where $ \Delta\one x_{\p}=\delta\one x_{\p} + \delta\one \lambda\,.$ The corresponding position on the perturbed FLRW spacetime is given by $  x^i_{\rm{FLRW}}(\lambda,{\n}) =   a(\bar{\eta})  x^i (\lambda,{\n}) $. Therefore, we need to re-map the conformal factor 
 \begin{eqnarray}\label{eq:re-mapa}
 a(\bar{\eta}) 
&=& a(\lambda_s)\left[1- \HH_s\delta\one \lambda-\frac{1}{2}\HH_s\left(\,{\delta}\two\lambda  
-2\Delta\one x_{\p}\partial_{\p} \delta\one \lambda -2\delta\one x_{\bot}^i\nabla_{\bot i} \delta\one \lambda
+\left( \frac{\HH'}{\HH_s}+ \HH_s\right) ( \delta\one \lambda)^2\right)\right]\,,
 \end{eqnarray} 
 Putting equation \eqref{eq:re-mapa} into $  x^i_{\rm{FLRW}}(\lambda,{\n}) =   a(\bar{\eta})  x^i (\lambda,{\n}) $ and using  equation \eqref{eq:trajectory3}  we find
 \begin{eqnarray}\label{eq:timemap4}
x^i_{\rm{FLRW}} (\lambda_s,{\n})&\approx& a(\lambda_s) r_s\bigg\{ {n}^i + \left[\frac{\delta\one x^i_{\bot}}{r_s} + \left[\frac{ \delta\one x_{\p}}{r_s} -\left(1- \frac{1}{ r_s\HH_s}\right){  \HH} \delta\one \lambda \right]n^i\right]
\\ \nonumber &&
+\frac{1}{2}\bigg[\frac{\delta\two x^i_{\bot}}{r_s} -\frac{2}{ r_s} \Delta\one x^j_{\bot} \nabla_{\bot j}\delta\one x^i_{\bot}
+\bigg( \frac{\delta\two x_{\p}}{r_s} - \frac{2}{r_s}\Delta\one x_{\p} \partial_{\p} \delta\one x_{\p}
\\ \nonumber &&
-\frac{2}{r_s}\delta\one x_{\bot}^i \nabla_{\bot i} \delta\one\delta x_{\p}
%\\ \nonumber &&
  - \left(1- \frac{1}{ r_s\HH_s}\right){  \HH}
 \left( \delta\two \lambda -2\Delta\one x_{\p} \partial_{\p} \delta\one \lambda   -2\delta\one x^j_{\bot}\nabla_{\bot j} \delta\one\lambda \right)\bigg)n^i\bigg]
\bigg\}\,,
\end{eqnarray}
At this point, we can define the deviation vector
\begin{eqnarray}
\frac{{{\xi}^i(\lambda_s,r{\n})}}{a(\lambda_s) r_s} &=& \frac{ {x^i_{\rm{FLRW}}(\lambda_s,r{\n})-\bar{x}^i_{\rm{FLRW}}(\lambda_s)}}{a(\lambda) r_s}=\frac{ {x^i_{\rm{FLRW}}(\lambda_s,r{\n})-\bar{x}^i_{\rm{FLRW}}(\lambda_s)}}{\bar{d}_{A}(\lambda)}\,.
\end{eqnarray}

%\subsection{Surface of constant redshift}

We measure the redshift  and angles and not the affine parameter. Therefore, we have to express the deviation vector in terms of the cosmological redshift. This implies that the contributions to the observed redshift due to Doppler effects, Sachs-Wolfe effect, integrated Sachs-Wolfe effects, etc are interpreted as displacements in the position of the source due to a local over-density. The distance to a source falling into a local over-density along the line of sight direction may appear shorter while sources moving away from the local over-density may appear stretched~\cite{Bolejko:2012uj}. 
The perturbation of the  observed redshift  is given by
 \begin{eqnarray}\label{eq:redshiftdef}
\left(1+z_{\rm{obs}}\right)
=\frac{E_s}{E_o}  =\frac{a({\bar{\eta}_{o}})}{a({\bar{\eta}_{s}})}\left[ 1+\delta\one z
+\frac{1}{2} \delta\two z\right]\,.
\end{eqnarray}
where $\delta\one z$ and $\delta\two z$ denotes the first and second order perturbations in redshift respectively. 
Using equation \eqref{eq:re-mapa}, we write equation \eqref{eq:redshiftdef} as 
    \begin{eqnarray}\label{eq:phya}
 \frac{1}{(1+{z}_{\rm{obs}} )}&=& \frac{a({\lambda_{s}})}{a({\lambda_{o}})}\bigg[1+\left(-\HH \delta\one \lambda - \delta\one z\right)
 \\ \nonumber &&
 +\frac{1}{2} \left(-\HH  \delta\two \lambda- \delta\two z +2 (\delta\one z)^2 +2\HH  \delta\one z \delta\one \lambda 
 + \left(\frac{\d \HH  }{\d \lambda_{s}} + \HH^2 \right)(\delta\one \lambda)^2\right)+\mathcal{O}(\epsilon^3\bigg]\,.
 \end{eqnarray}
Imposing that the redshift is  entirely due to Hubble flow(constant redshift surface)  implies that 
 \begin{eqnarray}  \label{eq;consistency1}
 \delta\one \lambda &=&-\frac{\delta\one z}{  \HH}, \\
   \label{eq;consistency2}
  \delta\two\lambda&=& -\frac{1}{\HH}\left[\delta\two z -(\delta\one z)^2 \left(1+ \frac{1}{\HH^2}\frac{\d\HH}{\d\lambda_{s}}\right)\right]\,.
 \end{eqnarray}
 Putting equations equations \eqref{eq;consistency1} and \eqref{eq;consistency2} in equation \eqref{eq:timemap4} gives
 \begin{eqnarray}\label{eq:deviationvecparallel1}
\frac{{{\xi}\one{\p}(z_s,r{\n})}}{\bar{d}_{A}}  &=&\frac{ \delta\one x_{\p}}{r_s} + \left(1- \frac{1}{ r_s\HH_s}\right)\delta\one z\,,
\\
\label{eq:deviationvecparallel2}
\frac{{{\xi}\two{\p}(z_s,r{\n})}}{\bar{d}_{A}}  &=&\frac{\delta\two x_{\p}}{r_s}-\frac{2}{r_s}\delta\one x_{\bot}^i \nabla_{\bot i} \delta\one\delta x_{\p}+\left(1- \frac{1}{ r_s\HH_s}\right) \left({\delta\two z} -{2}\frac{\delta\one z }{\HH}\partial\p \delta\one z -2\delta\one x^j_{\bot}\nabla_{\bot j} \delta\one z\right)\,.
\end{eqnarray}
The perturbed redshift in conformal Newtonian gauge  is given by
\begin{equation}\label{eq:perturbz}
\delta z=\left(\partial_{\|}{v_s}-  \partial_{\|}v_o\right) - \left( \Phi_s - \Phi_o\right)  -\int_{0}^{ r_s} \left({\Phi}' +  {\Psi}'\right){\d} r.
\end{equation}
Using equation \eqref{eq:perturbz} we find that the line of sight component of the  deviation  vector is given by
\begin{eqnarray}
\frac{{{\xi}\one{\p}(z_s,{\n})}}{a(z) r_s}  &=& \left(\partial_{\|}{v\one_s}-  \partial_{\|}v\one_o\right)  \left(1- \frac{1}{ r_s\HH_s}\right)\,,
\\
\frac{{{\xi}\two{\p}(z_s,{\n})}}{a(z) r_s} &=&\bigg[ \left(\partial_{\|}{v\two_s}-  \partial_{\|}v\two_o\right) -\frac{2}{\HH_s} \left(\partial_{\|}{v\one_s}-  \partial_{\|}v\one_o\right) \partial\p\partial_{\|}{v\one_s}
\\ \nonumber &&
-2\nabla_{\bot j}\partial_{\|}{v\one_s}\int^{ r_s}_0 {( r -  r_s)}\nabla_{\bot}^i(\Phi\one + \Psi\one) \d r\bigg]\left(1- \frac{1}{ r_s\HH_s}\right)\,.
\end{eqnarray}
Note  that we have expanded the peculiar velocity  up to second order $v = v\one + v\two/2$. 
The orthogonal component up to second order becomes
\begin{eqnarray}
\frac{\xi_{\bot }^{i} (z_s,{\n})}{a(z)r_s}= \frac{\delta\one x^i_{\bot}}{r_s} +\frac{1}{2}\left[ \frac{\delta\two x^i_{\bot}}{r_s}-\frac{2}{r_s} \Delta\one x^j_{\bot} \nabla_{\bot j}\delta\one x^i_{\bot}\right]  \,,
\end{eqnarray}
where
\begin{eqnarray}
\delta\one x^i_{\bot}(z_s,{\n}) &
=&\int^{ r_s}_0 {( r -  r_s)}\nabla_{\bot}^i(\Phi\one + \Psi\one) \d r\,,
%\\   %\nonumber 
%\delta\two x^i_{\bot}(\bar{\eta}_s,{\n})&=&
%\int_{0}^{ r_s} ({ r} -  r_s)\nabla_{\bot}^i(\Phi\two + \Psi\two)\d r 
%\\   \nonumber &&
%+8\int_{0}^{ r_s}{\d} r\int_{0}^{ r} \nabla^j_{\bot }\Phi\one({ r'} {\n})\d{ r' }\int_{0}^{ r} \frac{({ r''} -  r){ r''}}{ r} \nabla_{\bot j}\nabla^i_{\bot }\Phi\one({ r''}{\n}) \d{ r''}\,. \qquad
\end{eqnarray}
The weak gravitational lensing  convergence  at first order is given by
\begin{eqnarray}
  \kappa\one(z_s,{\n}) &=&- \frac{1}{2}\nabla_{\Omega A}\xi_{\bot }^{A}= \frac{1}{2}\int^{ r_s}_0 \frac{( r -  r_s) r}{ r_s}\nabla_{\bot}^2(\Phi\one + \Psi\one) \d r \,,
 % \\
 %  \kappa\two &=&\frac{1}{2}\nabla_{\bot A} {\Delta x\two_{\bot }}^{A}
\end{eqnarray}
where $\nabla_{\bot}^2 = \nabla_{\bot i} \nabla_{\bot} ^i $,  we will not need the explicit form of $   \kappa\two $ for our calculation, hence no need to give it here.
The twist vanishes $\omega_{ij}= 0$ at linear order. The shear at first order is given by
\begin{eqnarray}
\gamma\one_{ij}(z_s,{\n})=\nabla_{\Omega \<i}\Delta\one x_{\bot j\>} = \int^{ r_s}_0\frac{(\tilde{ r} -  r_s)\tilde{ r}}{ r_s}\nabla_{\bot \<i}\nabla_{\bot j\>}(\Phi\one + \Psi\one)\d\tilde{ r}\,.
\end{eqnarray}

\section{Area distance from the null focusing  equations}\label{sec:Sachsequation}

The focusing equation for the area distance is given by~\cite{Umeh:2012pn}
\begin{eqnarray}\label{eq:areadistanceqn}
\frac{\d^2 d_A}{\d\lambda^2}=-\left[\frac{1}{2}  R_{ab} k^a k^b+{\Sigma_{ab}}\Sigma^{ab}\right] {d}_A\,,
\end{eqnarray}
where $R_{ab}$ is the Ricci tensor, $\Sigma_{ab}$ is the null shear associated with $k^a$.  
Equation \eqref{eq:areadistanceqn} may be solved perturbatively with the following  initial conditions
 \begin{eqnarray}
 \delta^{n} d_A\bigg|_o =0~~~\rm{and}~~~
\frac{\d\delta^n d_A}{\d\lambda}\bigg|_o = - \delta^n E_o\,,
 \end{eqnarray}
 where the perturbations in the photon energy at the observer location is given by
 \begin{eqnarray}
  \delta E_o &=&   \Phi\one_o \,, \qquad \qquad 
   \delta^2 E_o = \Phi\two_o 
 -{\Phi\one}^2_o\,.
 \end{eqnarray}
 Using these initial conditions, the solution to equation \eqref{eq:areadistanceqn} becomes
 \begin{eqnarray}\label{DAeqn1soln}
 \frac{\delta\one d_A }{\bar d_A}&=& -    \int_{\lambda_o}^{\lambda_s}  {\d}\lambda\frac{ {(\lambda_s-\lambda) ( \lambda_o- \lambda)}}{(\lambda_o-\lambda_s)} \nabla^2_{\bot}\Phi\one \, ,
\\ \nonumber
  \frac{\delta\two d_A }{\bar d_A}&=& 
    -2 \int_{\lambda_o}^{\lambda_s} {\d}\lambda\frac{(\lambda_s-\lambda)}{(\lambda_o-\lambda_s)}\nabla_{\bot}^2\Phi\one\int_{\lambda_o}^{\lambda} {\d\tilde{\lambda}}(\lambda-\tilde{\lambda})(\lambda_o-\tilde{\lambda})\nabla_{\bot}^2\Phi\one(\tilde{\lambda})
        \\ \nonumber&&
    -8 \int_{\lambda_o}^{\lambda_s} {\d}\lambda\frac{(\lambda_s-\lambda)}{(\lambda_o-\lambda_s)}\int^{\lambda}_{\lambda_s} {\d}{\tilde{\lambda}}\nabla_{\bot i}\Phi\one(\tilde{\lambda})\int^{\lambda}_{\lambda_o} {\d\tilde{\lambda}}\frac{(\lambda_o-\tilde{\lambda})^2}{(\lambda_o-\lambda)}\nabla_{\bot}^i\nabla^2\Phi\one(\tilde{\lambda})
       \\ \nonumber&&
  -4 \int_{\lambda_o}^{\lambda_s} {\d}\lambda\frac{(\lambda_s-\lambda)}{(\lambda_o-\lambda_s)}\nabla_{\bot i}
  \Phi\one\int^{\lambda}_{\lambda_o}{\d}{\tilde{\lambda}}\frac{(\lambda_s-\tilde{\lambda})(\lambda_o-\tilde{\lambda})^2}{(\lambda_o-\lambda)}\nabla_{\bot}^i\nabla_{\bot}^2\Phi\one(\tilde{\lambda})
       \\ &&
  -4 \int_{\lambda_o}^{\lambda_s} {\d}\lambda\frac{(\lambda_s-\lambda)(\lambda_o-\lambda)}{(\lambda_o-\lambda_s)}\int_{\lambda_o}^{\lambda} {\d}{\tilde{\lambda}}\nabla_{\bot\<i}\nabla_{\bot j\>}\Phi\one(\tilde{\lambda})\int_{\lambda_o}^{\lambda} {\d}\tilde{\lambda}\nabla_{\bot}^{\<i}\nabla_{\bot}^{ j\>}\Phi\one(\tilde{\lambda})
 \,.
 \end{eqnarray}
 where we have focused on the dominant terms only. 
On the constant redshift surface, the area distance becomes\cite{Umeh:2014ana}
   \begin{eqnarray}\nonumber
{d}_A(z,{\n})&=&\bar{d}_{A}(z_s)
\bigg\{1+\left[\frac{\delta\one d_A}{\bar{d}_A}+\left(1- \frac{1}{\HH_s  r}\right)\delta\one z\right]
%\\ \nonumber &&
 +
 \frac{1}{2}\bigg[\frac{ \delta\two d_A}{\bar{d}_A} 
 + 2\frac{\delta\one d_A}{\bar{d}_A}\delta \one z\left(1- \frac{1}{\HH_s  r}\right)
 +  \frac{(\delta\one z)^2}{\HH r_s}\left(
 \frac{\HH'_s}{\HH^2}- 1\right)
 \\ \nonumber&&
    + \frac{2}{\HH_s}\delta\one z \left(\frac{\delta\one d_A}{\bar{d}_A}\right)' + 2\Delta\one x_{\p} \partial_{\p} \left(\frac{\delta\one d_A}{\bar{d}_A}\right) 
 + 2\Delta\one x_{\bot}^i\nabla_{\bot i} \left(\frac{\delta\one d_A}{\bar{d}_A}\right)
\\  &&
 +2\left(\delta\two z+\frac{1}{\HH_s}\delta\one z \delta\one z' - \Delta\one x_{\p} \partial_{\p} \delta\one z - \Delta\one x_{\bot}^i\nabla_{\bot i} \delta\one z\right)\left(1- \frac{1}{\HH_s  r}\right)
 % \right.\\ \nonumber&&\left.
 \bigg]
 \bigg\}.\label{eq:DAgen2}
 \end{eqnarray}
The leading order part is given by
\begin{eqnarray}
%{{d}_A}(z,{\n})&\approx&\bar{d}_{A}(z_s)
%\bigg\{1+{\frac{{{\xi}{\p}(z_s,{\n})}}{a(z) r_s}  } + {\frac{ \Delta d_A}{\bar{d}_A} }\bigg\}\,,\\
{{d}_A}(z,{\n})&\approx&\bar{d}_{A}(z_s)\bigg\{1+\left(1- \frac{1}{\HH_s  r}\right)\delta\one z + \frac{1}{2} {\frac{ \delta d_A}{\bar{d}_A} }
%\\ \nonumber &&
+
\frac{1}{2}\left(\delta\two z-2\Delta\one x_{\p} \partial_{\p} \delta\one z -2 \Delta\one x_{\bot}^i\nabla_{\bot i} \delta\one z\right)\left(1- \frac{1}{\HH_s  r}\right)
\bigg\}\,.\qquad
\end{eqnarray}
\iffalse
where $\averageA{{ \Delta d_A}/{\bar{d}_A} }$ denotes the non-vanishing shear contribution
\begin{equation}
\averageA{\frac{ \Delta d_A}{\bar{d}_A} }\sim     -4 \int_{\lambda_o}^{\lambda_s} {\d}\lambda\frac{(\lambda_s-\lambda)(\lambda_o-\lambda)}{(\lambda_o-\lambda_s)}\int_{\lambda_o}^{\lambda} {\d}{\tilde{\lambda}}\nabla_{\bot\<i}\nabla_{\bot j\>}\Phi\one(\tilde{\lambda})\int_{\lambda_o}^{\lambda} {\d}\tilde{\lambda}\nabla_{\bot}^{\<i}\nabla_{\bot}^{ j\>}\Phi\one(\tilde{\lambda})\,.
\end{equation}
 \fi

 \section{Spherical Harmonic decomposition of the key term}\label{sec:harmonicsdecomp}

 We use the Poisson equations  to express the velocity potential  in terms of the matter density contrast $\delta_m$ 
 \begin{eqnarray}
 v({\k},\eta) &=& \frac{\HH}{k^2}f(\eta)\delta_m({\k},\eta)\,.\label{eq:continuityeqn}
 %\\
%  \Phi({\k},\eta)& =& -\frac{3}{2}\Omega_m(z) \left(\frac{\HH}{k}\right)^2\delta_m({\k},\eta)\,.
  \label{eq:Poissonequn}
 \end{eqnarray}
 The most important term in our analysis is 
 \begin{eqnarray}
   \averageA{\frac{\xi{\p}}{\bar{d}_A} }^{A}&\approx& - \left(1 - \frac{1}{ r_{s} \HH_{s}}\right) \averageA{
\frac{1}{\HH_{s}} \left(\partial_{\|}{v\one_s}-  \partial_{\|}v\one_o\right) \partial^2\p{v\one_s}}\,,
\\
&=& - \left(1 - \frac{1}{ r_{s} \HH_{s}}\right) \frac{1}{\HH_{s}}\bigg[\averageA{
 \partial_{\|}{v\one_s} \partial^2\p{v\one_s}} -\averageA{
 \partial_{\|}{v\one_o} \partial^2\p{v\one_s}} \bigg] \,.\label{eq:AveragexiA}
 \end{eqnarray}
% In the second equality, we dropped the peculiar velocity evaluated at the observer position. When expanded in spherical harmonics, it can only contribute at $\ell=1$. Dropping it at this point implies that we can only sum the multipoles from $\ell =2$.
% Now, we describe how we perform the full-sky spherical harmonics of decomposition of equation \eqref{eq:AveragexiA}. 
Expanding each of the terms in $\averageA{\partial_{||} v\one_{s} \partial^2_{||} v\one_{s}} $ in Fourier space we find
\begin{eqnarray}\label{eq:postborn1}
\averageA{\partial_{||} v\one_{s} \partial^2_{||} v\one_{s}} &=& - i\left(D(z)\HH(z)f(z)\right)^2 \int {\d {\n} } \int \frac{\d^3 k_{1}}{(2\pi)^3}\int \frac{\d^3 k_{2}}{(2\pi)^3}
\frac{1}{2}
\left[ \frac{ \mu_{1} \mu^2_{2}}{k_{1}}  + \frac{\mu_{2} \mu^2_{1}}{k_{2}} \right]
%\\ \nonumber && \times
 e^{i k_{1}\cdot {\x}}e^{i k_{2}\cdot {\x}} \delta_{m}({\k}_{1})\delta_{m}({\k}_{2})\,,
\end{eqnarray}
where $\mu_{i} = \hat{\k}_{i} \cdot {\n}$ and $i = 1,2$. We made use of equation \eqref{eq:continuityeqn} to express the velocity field in terms of the density field. 
If we take the ensemble average of  $\overline{\averageA{\partial_{||} v\one_{s} \partial^2_{||} v\one_{s}} }$ at this stage using
\begin{eqnarray}\label{eq:powerspec}
\overline{\delta\one_{{\rm{m}}}({\k}_{1})\delta\one_{{\rm{m}}}({\k}_{2}) } = (2\pi)^3 P_{m}(k_1)\delta^{(D)}\left({\k}_{1} +{\k}_{2}\right) %= (2\pi)^3 P_{m}(k_1)\delta^{(D)}\left({\k}_{1} +{\k}_{2}\right)
\end{eqnarray}
equation \eqref{eq:postborn1} vanishes. %This is the well-known  flat-sky approximation. 
We go beyond this to  decompose $ e^{i k_{1}\cdot {\x}}$ in a linear combination of spherical waves,
\begin{equation}
  e^{i{\k}\cdot{\x}}=4\pi \sum_{\ell m}i^{\ell} j_{\ell} (kr) Y^{\ast}_{\ell m}({\n}) Y_{\ell m}(\hat{\k})\,.
\end{equation}
It allows to expand the angles in spherical harmonics are well
 \begin{eqnarray}
 i\mu e^{ikr}  &=& n^i\partial_{i} e^{ikr}= 4\pi \sum_{\ell m} i^\ell j_{\ell}'(kr) Y^{\ast}_{\ell m}({\n}) Y_{\ell m}(\hat{\k})\,,
 \\ 
 -\mu^2 e^{ikr}&=& \partial^2_{\p}e^{i kr} =4\pi \sum_{\ell m} i^\ell j_{\ell}''(kr) Y^{\ast}_{\ell m}({\n}) Y_{\ell m}(\hat{\k})\,.
 \end{eqnarray}
 Substituting this into equation \eqref{eq:postborn1} leads to 
 \begin{eqnarray}\label{eq:postborn2}
\averageA{\partial_{||} v\one_{s} \partial^2_{||} v\one_{s}} &=& (4\pi)^2\left(D(z)\HH(z)f(z)\right)^2\sum_{\ell_1 m_1}\sum_{\ell_2 m_2}i^{(\ell_1+ \ell_2)} 
\int {\d {\n }} \int \frac{\d^3 k_{1}}{(2\pi)^3}\int \frac{\d^3 k_{2}}{(2\pi)^3}  \delta_{m}({\k}_{1})\delta_{m}({\k}_{2})\
\\ \nonumber && \times
\frac{1}{2}
\left[ \frac{ j'_{\ell}(k_1r)j''_{\ell}(k_2r)}{k_{1}}  + \frac{j'_{\ell}(k_2r) j''_{\ell}(k_1r)}{k_{2}} \right]
Y^{\ast}_{\ell_1 m_1}({\n})
Y^{\ast}_{\ell_2 m_2}({\n})
Y_{\ell_1 m_1}(\hat{\k}_1)Y_{\ell_2 m_2}(\hat{\k}'_{2})\,,
\end{eqnarray}
At this point. we can take the ensemble average using equation \eqref{eq:powerspec}
 \begin{eqnarray}\label{eq:postborn3}
\overline{\averageA{\partial_{||} v\one_{s} \partial^2_{||} v\one_{s}} }&=& (4\pi)^2\left(D(z)\HH(z)f(z)\right)^2\sum_{\ell_1 m_1}\sum_{\ell_2 m_2}i^{(\ell_1+ \ell_2)} (-1)^{\ell_2}\int  {\d {\n} } \int \frac{\d^3 k_{1}}{(2\pi)^3} P_{m}(k_1)\
\\ \nonumber && \times
\left[ \frac{ j'_{\ell}(k_1r)j''_{\ell}(k_1r)}{k_{1}}   \right]
Y^{\ast}_{\ell_1 m_1}({\n})
Y^{\ast}_{\ell_2 m_2}({\n})
Y_{\ell_1 m_1}(\hat{\k}_1)Y_{\ell_2 m_2}(\hat{\k}_{1})\,.
\end{eqnarray}
We have made use of the Parity symmetry of the spherical harmonics: ${ Y_{\ell m}(-\hat{\k} )=(-1)^{\ell }Y_{\ell m}( \hat{\k} )}$. We can use the spherical harmonics addition rule to switch the position of the conjugation 
\begin{equation}
\mathcal{L}_{\ell}({\n}\cdot\hat{\k}_{1}) =  \frac{4\pi}{(2\ell_1+1)}\sum_{m_{1}=-\ell_1}^{\ell_1}Y_{\ell_1 m_1}^{\ast}({\n})Y_{\ell_1 m_1}(\hat{\k}_1) = 
 \frac{4\pi}{(2\ell_1+1)}\sum_{m_{1}=-\ell_1}^{\ell_1}Y_{\ell_1 m_1}({\n})Y^{\ast}_{\ell_1 m_1}(\hat{\k}_1)\,,
\end{equation}
where $\mathcal{L}_{\ell}$ is the Legendre polynomial.  This allows to  perform the angular k-integral using the orthogonality condition
\begin{eqnarray}
\int d \Omega_{k_2}  Y_{\ell_2 m_2} (\hat{\k}_1)Y^*_{\ell_{1} m_{1}} (\hat{\k}_1) = \delta_{\ell_2 \ell_{1}} \delta_{m_2 m_{1}}\,.
\end{eqnarray}
Then  we can perform $\ell_{2}$ and $m_{2}$ sums. Using the addition theorem, we sum over the remaining $m_{1}$ using
\begin{eqnarray}
{ \sum _{m=-\ell }^{\ell }Y_{\ell m}^{*}({\n} )\,Y_{\ell m}({\n} )={\frac {2\ell +1}{4\pi }}}
\end{eqnarray}
leading to 
\begin{eqnarray}
\overline{\averageA{\partial_{||} v\one_{s} \partial^2_{||} v\one_{s}} }&=& (D(z)\HH(z)f(z))^2 \sum_{\ell = 0}^{\ell_{\rm{max}}} (2\ell+1)\int \frac{\d k}{2\pi^2} k P_{m}(k)  j'_{\ell}(kr) j''_{\ell}(kr)\,.
\end{eqnarray}
In the body of the paper, we dropped the overbar(ensemble average) to reduce clutter. 
%For terms involving angular derivatives, we showed in  \cite{Umeh:2021} that the Total Angular Momentum approach introduced in the CMB analysis~\cite{Hu:1997hp} is best suited for its full-sky spherical harmonics decomposition.

Similarly, 
\begin{eqnarray}
\overline{\averageA{\partial_{\|}{v\one_o} \partial^2\p{v\one_s}} }&=& (D(z)\HH(z)f(z))(D(z_o)\HH(z_o) f(z_o)) \sum_{\ell = 0}^{\ell_{\rm{max}}} (2\ell+1)\int \frac{\d k}{2\pi^2} k P_{m}(k)  j'_{\ell}(kr_o) j''_{\ell}(kr)\,.
\end{eqnarray}

We use this correspondence for efficiency (see figure \ref{fig:shpericalDecomp})
\begin{eqnarray}\label{eq:correspond}
\overline{\averageA{\partial_{||} v\one_{s} \partial^2_{||} v\one_{s}} } -\overline{ \averageA{\partial_{\|}{v\one_o} \partial^2\p{v\one_s}}}&\approx&\overline{\averageA{\partial_{||} v\one_{s} \partial^2_{||} v\one_{s}} } \,, \qquad{\rm{with}}\qquad \ell_{\rm{min}} = 2.
\\
 &=&(D(z)\HH(z)f(z))^2 \sum_{\ell = 2}^{\ell_{\rm{max}}} (2\ell+1)\int \frac{\d k}{2\pi^2} k P_{m}(k)  j'_{\ell}(kr) j''_{\ell}(kr)\,.
 %\\
%&=& { (D(z)\HH(z)f(z))(D(z_o)\HH(z_o) f(z_o)) \sum_{\ell = 0}^{\ell_{\rm{max}}} (2\ell+1)\int \frac{\d k}{2\pi^2} k P_{m}(k)  j'_{\ell}(kr_o) j''_{\ell}(kr)}\,.
\qquad
\end{eqnarray}
Using the following relationships for the derivatives of th spherical Bessel functions
\begin{eqnarray}\label{eq:jprA1}
j'_{\ell}(x) &=& -j_{\ell +1} +\frac{\ell}{x} j_{\ell}(x)\qquad {\rm{for}}\qquad \ell = 0,1,2\cdots
\\ \label{eq:jprA2}
j''_{\ell}(x)&=&\frac{1}{x^2} \left[\left( \ell^2- \ell-x^2\right)j_{\ell}(x)+ 2 x j_{\ell+1}(x)\right]\,,\qquad {\rm{for}}\qquad \ell = 0,1,2\cdots
%\\
%j'''_{\ell}(x) &=&  \frac{1}{x^2}\left[ \left( \ell^2 - \ell -x^2\right)  j'_{\ell}(x)  - 2 x j_{\ell} (x) + 2 x j'_{\ell+1}(x) + 2j_{\ell+1}(x)\right] - \frac{2}{x^3}
% \ell = 0,1,2\cdots
\end{eqnarray}
%with $j'_{0}(x) = -j_{1}(x)$.
We show the  correspondence in equation \eqref{eq:correspond} in figure \ref{fig:shpericalDecomp}. 

     \begin{figure}[h]
%\centering 
\includegraphics[width=100mm,height=75mm] {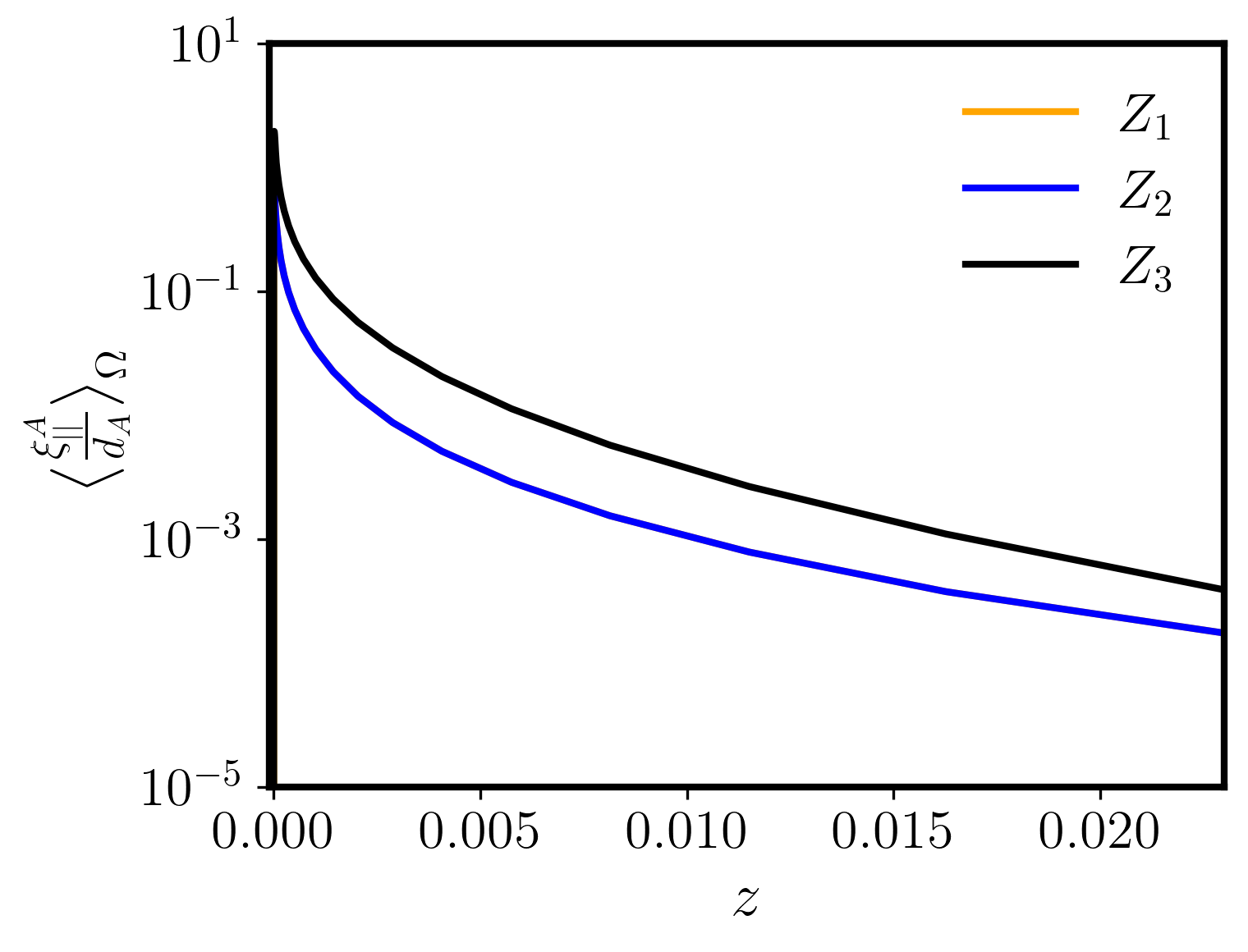}
\caption{\label{fig:shpericalDecomp}  {  We show the correspondence employed to calculate $\averageA{{\xi{\p}}/{\bar{d}_A} }^{A}$ in the body of the paper. The definitions of the legend is given in equations \eqref{eq:Z1}-\eqref{eq:Z3}. We summed from $\ell=2$ to $\ell =20$ for $Z_1$ and $Z_2$ establishing the approximation used in the body of the paper that $Z_1=Z_2$ with the contribution of  $\ell=0$ and $\ell=1$ removed.  $Z_3$ gives the total contribution for an observer positioned in the barycenter of our local group. }
 %\red{ I am running computation for higher $\ell_{\rm{max}}$ to show that they agree exactly. The kink at about $z=0.6$ disappears as $\ell_{\rm{max}}$ is inceased}
}
\end{figure}
{
\begin{eqnarray}\label{eq:Z1}
Z_1&=&- \left(1 - \frac{1}{ r_{s} \HH_{s}}\right) \bigg[ \HH(z_s)( f(z_s)D(z_s))^2 \sum_{\ell =2}^{\ell_{\rm{max}}} (2\ell+1)\int \frac{d k_1}{2\pi^2} k_{1}P_{m}(k_1) j'_{\ell}(k_1  r_s)j''_{\ell}(k_1  r_s)
\\ \nonumber &&
-(D(z)\HH(z)f(z))(D(z_o)\HH(z_o) f(z_o)) \sum_{\ell = 2}^{\ell_{\rm{max}}} (2\ell+1)\int \frac{\d k}{2\pi^2} k P_{m}(k)  j'_{\ell}(kr_o) j''_{\ell}(kr)\bigg]\,,
\\
Z_2&=&- \left(1 - \frac{1}{ r_{s} \HH_{s}}\right)  \HH(z_s)( f(z_s)D(z_s))^2 \sum_{\ell =2}^{\ell_{\rm{max}}} (2\ell+1)\int \frac{d k_1}{2\pi^2} k_{1}P_{m}(k_1) j'_{\ell}(k_1  r_s)j''_{\ell}(k_1  r_s)\,,
\label{eq:Z2}
\\
Z_3 &=& \left(1 - \frac{1}{ r_{s} \HH_{s}}\right) (D(z)\HH(z)f(z))(D(z_o)\HH(z_o) f(z_o)) \sum_{\ell = 0}^{\ell_{\rm{max}}} (2\ell+1)\int \frac{\d k}{2\pi^2} k P_{m}(k)  j'_{\ell}(kr_o) j''_{\ell}(kr)\,..
\label{eq:Z3}
\end{eqnarray}
}

\newpage
%For $r_o=0$
%\bibliographystyle{$HOME/Dropbox/UWC_papers/Effective_fnl/Biassecondorder/JHEP}%{JHEP}%
%\bibliographystyle{apsrev4-1}
%\bibliography{cosmoref.bib}

%\bibliographystyle{JHEP}
%\bibliography{$HOME/Dropbox/UWC_papers/q-dipole/draft/cosmoref.bib}

\providecommand{\href}[2]{#2}\begingroup\raggedright\endgroup

\end{document}